\documentclass[12pt]{article}
\usepackage{a4wide}
\usepackage{graphicx}
\usepackage{amssymb}

\newcommand{\be}{\begin{equation}}
\newcommand{\ee}{\end{equation}}
\newcommand{\ba}{\begin{eqnarray}}
\newcommand{\ea}{\end{eqnarray}}
\newcommand{\bi}{\begin{itemize}}
\newcommand{\ei}{\end{itemize}}
\newcommand{\lag}{\mathcal{L}}
\renewcommand{\d}[1]{\frac{d}{d#1}}
\newcommand{\dd}[1]{\frac{d^2}{d#1^2}}
\newcommand{\dsp}{\displaystyle}

\renewcommand{\theequation}{\arabic{section}.\arabic{equation}}
\begin{document}
\begin{titlepage}
\begin{flushright}
LU TP 03-32\\
hep-ph/0307044\\
revised October 2003
\end{flushright}
\vfill
\begin{center}
{\Large\bf Scalar Form Factors in SU(3) Chiral Perturbation Theory}\\
\vfill
{\bf Johan Bijnens and Pierre Dhonte}\\[1cm]
{Department of Theoretical Physics, Lund University\\
S\"olvegatan 14A, S 22362 Lund, Sweden}
\end{center}
\vfill
\begin{abstract}
The scalar form factors for the pions and kaons are calculated in SU(3) Chiral
Perturbation Theory at order $p^6$, in the isospin limit $m_u=m_d=\hat{m}$.
We find in general sizable corrections of  $\mathcal{O}(p^6)$.
We use our results to obtain information on the $1/N_c$ suppressed
low energy constants $L_4^r$ and $L_6^r$ as well as on two
$\mathcal{O}(p^6)$ low energy constants. We present some numerical results
for masses and decay constants as well.

We derive a relation for the scalar form factors analogous to
Sirlin's relation for vector form factors.
\end{abstract}
 \vfill
 \end{titlepage}
\section{Introduction}
\label{introduction}

Chiral Perturbation Theory (ChPT) in its effective Lagrangian form
was introduced by Weinberg \cite{Weinberg}
and developed into its standard form, for both two and three
light flavours (referred to as $SU(2)$ and $SU(3)$)
formalisms, up to order $p^4$, by Gasser and
Leutwyler \cite{GL1,GL2}.
The underlying assumption of ChPT
is that the chiral $SU(n)_L \times SU(n)_R$ symmetry present
in the QCD Lagrangian is spontaneously broken to its vector
subgroup, $SU(n)_L \times SU(n)_R \rightarrow SU(n)_V$.
This breaking is due to the finite quark condensates\footnote{See below
for the alternative view of generalized ChPT.}
\be 
\label{qcondens}
\langle 0|\bar{q}q|0\rangle \neq 0\mbox{, \quad for $q$=$u$,$d$ and $s$.}
\ee
The quark and gluon degrees of freedom are integrated out and
replaced by pseudo-scalar fields representing the lightest pseudo-scalar
mesons triplet ($SU(2)$) or octet ($SU(3)$), these are the Goldstone
bosons resulting from the spontaneous chiral symmetry breaking.
Introducing non-zero quark masses renders the chiral symmetry only
approximate. This results in masses for the mesons.

ChPT is thus an effective theory of QCD, perturbatively expanded in powers of
the relevant momentum and the quark masses, and is non-renormalizable. 
Loop corrections produce ultra-violet divergences which require the
introduction of new counterterms and associated low energy constants (LEC) at
each order in the parameters expansion. 
The number of new operators is, however, finite at each order, allowing
ChPT to be predictive. Recent introductions to ChPT can be
found in Ref.~\cite{chptlectures}.

A basic set of calculations in $SU(3)$ ChPT at order $p^4$ was performed
in \cite{GL2,GL3} and used to determine the
LEC at order $p^4$, the $L_i^r$, to that order.
$L_4^r$ and $L_6^r$ were assumed to be negligible there
using the large $N_c$ limit of QCD \cite{largeNc}. This assumption
was studied at order $p^4$ in \cite{Moussallam1} by considering
scalar form factors and a scalar two-point function.

The state of ChPT in the meson sector now is such that many calculations
have been pushed to the next order, $p^6$ or next-to-next-to-leading
(NNLO) or two-loop order. 
There are two main reasons to explore the two-loop region of 
$SU(3)$ Chiral Theory. First, the accuracy of experimental results
has improved so this accuracy is now routinely needed
and we need to determine all LECs to the required precision.
Second, spontaneous chiral symmetry breaking of QCD does not have to be driven
by the quark condensates, Eq. (\ref{qcondens}). An alternative bookkeeping
method in the Lagrangian expansion might be necessary, leading to a Generalized
ChPT (GChPT), see Ref.~\cite{Stern1} and references therein.
The combination of the two-loop calculation of $\pi\pi$-scattering
in $SU(2)$ ChPT \cite{pipi1,pipi2}, together with the dispersive Roy equation
analysis \cite{ACGL}, lead to sharp predictions for the
$\pi\pi$ scattering lengths \cite{CGL1} confirmed by the E865 experiment
\cite{E865} leading to the conclusion that $SU(2)$ ChPT is indeed driven
by the quark condensate \cite{CGL2}. See also \cite{Stern1,Stern2,Stern3}
for relevant discussions. In the case of $SU(3)$ ChPT the possibility
of a small quark condensate is not ruled
out \cite{Moussallam1,Moussallam2,paramagnetic}.
The pion scalar form factor has been evaluated to order $p^6$
in $SU(2)$ ChPT in \cite{BCT} after earlier dispersive work \cite{GM}.

The two-loop calculations in $SU(3)$ ChPT
of the masses \cite{ABT1,GK} indicated the possibility of sizable
NNLO corrections. This behaviour persisted when the calculations
also included $K_{\ell4}$ \cite{ABT2,ABT3}
such that the $L_i^r$ could be determined
with $p^6$ accuracy under similar assumptions as the $p^4$ determination
\cite{GL2}. In \cite{ABT3} a search was done for possible values of
$L_4^r$ and $L_6^r$ that would lead to acceptable convergence of all
quantities considered there. In \cite{ABT4} effects of isospin violation
in the masses were taken into account also to order $p^6$ as well
as well as the new E865 $K_{\ell4}$ data of \cite{E865}.

This paper consists of two major parts. In the
first part we calculate the scalar form factors of pions and kaons
to order $p^6$ in $SU(3)$ ChPT and show some numerical results for the sizes
of the various contributions to the different form factors.
We therefore first present the
framework of $SU(3)$ ChPT in the isospin limit and 
including external scalar fields in Section~\ref{lagrangian} and
sketch the calculation of the scalar form factors in Section~\ref{FF}.
The analytical results of this calculation are given in 
Section~\ref{C_contribs} and App.~\ref{Appresults}.
In Section~\ref{plots} we present a series of plots allowing the reader
to judge the importance of the different parts of the calculations.
One result of more general interest is that
the curvature of the scalar form factor
in $K_{\ell3}$ decays \cite{BT2}
can in principle be determined from the quantities considered here.
Unfortunately as discussed at the end of Sect.~\ref{c12c13} the numerical
accuracy obtainable is rather low. 

Those results allow us then to proceed to the second major part of this paper.
A study of the form factors allowing to relax some of the assumptions made in
the previous $L_i^r$ determinations and the effects of these assumptions
on the scalar form factors. 
We made use of dispersive analysis, Section~\ref{experiments}, to extract
experimental information. This allows us to obtain extra
numerical results for some of the LEC's of the theory,
Sections~\ref{L4L6} and \ref{c12c13}. We also
present some updated results on masses and decay constants in
Sect.~\ref{Masses}.

A third result is a relation between the scalar form factors
that has only second order corrections in quark mass differences.
This relation, Eq.~(\ref{eq:relation}) is proven in general in
App.~\ref{app:relation} and is valid for all $t$ where an expansion in mass
differences is valid. It is the analog of the relation derived
by Sirlin for the vector form factors~\cite{Sirlin}.

\section{The ChPT Lagrangian}
\label{lagrangian}

This section contains a very short overview of ChPT and serves also
to define our conventions. More detailed introductions can be found in
\cite{chptlectures}. The Lagrangians of ChPT are ordered in the $p$
counting, powers of momenta ($\mathcal{O}(p)$), quark masses or
external scalar or pseudo scalar fields ($\mathcal{O}(p^2)$)
and external vector or axial-vector fields ($\mathcal{O}(p)$).
We keep to the three flavour case here.

The Lagrangian for the strong and semi leptonic mesonic sector
to NNLO can be written as
\be
\lag = \lag_2+\lag_4+\lag_6\,,
\ee
where the subscript refers to the chiral order.
The lowest order Lagrangian
\be
\label{L2}
{\cal L}_{2} = \frac{F_0^2}{4} \langle u_\mu u^\mu + \chi_+ \rangle \,.
\ee
The mesonic fields enter via
\be
u = \exp\left(\frac{i M}{F_0 \sqrt{2}}\right)\,,\quad
M =\left(\begin{array}{ccc}
\frac{1}{\sqrt{2}}\pi^0+\frac{1}{\sqrt{6}}\eta & \pi^+ & K^+\\
\pi^- & \frac{-1}{\sqrt{2}}\pi^0+\frac{1}{\sqrt{6}}\eta & K^0\\
K^- & \overline{K^0} & \frac{-2}{\sqrt{6}}\eta
         \end{array}\right)\,
\ee
and the quantity $u_\mu$ introduces the external vector ($v_\mu$)
and axial-vector ($a_\mu$)
currents
\be
\label{covariant}
u_\mu = i 
(u^\dagger \partial_\mu u - \partial_\mu u u^\dagger 
 -i u^\dagger r_\mu u + i u l_\mu u^\dagger)\,,
\quad
l_\mu(r_\mu) = v_\mu -(+) a_\mu\,, 
\ee
while the scalar ($s$) and pseudo scalar ($p$) currents are contained in
\be
\chi_\pm = u^\dagger \chi u^\dagger \pm u \chi^\dagger u\,,\quad
\chi = 2 B_0\left(s+ip\right)\,.
\ee
The $p^4$ or NLO Lagrangian $\lag_4$ was introduced in Ref.~\cite{GL2}
and reads
\ba
\label{L4}
{\cal L}_{4}&=&
\hspace{0.34cm} L_1 \langle u_\mu u^\mu \rangle^2
+L_2 \langle u_\mu u_\nu  \rangle
     \langle u^\mu u^\nu  \rangle  \nonumber\\&& 
+L_3 \langle u^\mu u_\mu u^\nu u_\nu \rangle 
+L_4 \langle u^\mu  u_\mu  \rangle \langle \chi_+ \rangle \nonumber\\&&
+L_5 \langle u^\mu u_\mu \chi_+ \rangle
+L_6 \langle \chi_+ \rangle^2 \nonumber\\&&
+L_7 \langle \chi_- \rangle^2
+L_8 \langle \chi_+ \chi_- \rangle \nonumber\\&&
\hspace{-0.14cm}
-i L_9 \langle F^R_{\mu\nu} u u^\mu u^\nu u^\dagger 
             + F^L_{\mu\nu} u^\dagger u^\mu u^\nu u \rangle 
+L_{10} \langle   F^R_{\mu\nu}  U F^{L\mu\nu} U^\dagger 
\rangle 
\nonumber\\&&
+H_1 \langle F_{\mu\nu}^RF^{\mu\nu R} + F_{\mu\nu}^LF^{\mu\nu L} \rangle
+H_2 \langle \chi_+^2 - \chi_-^2 \rangle /4 
\,.
\ea
 The $L_9$ and $L_{10}$ terms introduce also the field strength tensor
\be
F_{\mu\nu}^{L(R)} = \partial_\mu l(r)_\nu -\partial_\nu l(r)_\mu -i 
\left[ l(r)_\mu , l(r)_\nu \right]\,.
\ee
The two terms proportional to
$H_1$ and $H_2$ are contact terms and do not enter physical amplitudes.
In the present case, we keep only the relevant scalar current interactions
from which we extract and separate the quark masses contribution:
\be
s = 
\left(\begin{array}{ccc}\hat{m} &  & \\ & \hat{m} & \\ & & m_s
\end{array}\right)\,+\,\tilde{s}\,,
\quad
p = l_\mu = r_\mu = 0\,.
\ee
We quote the schematic form of the NNLO Lagrangian in the $SU(3)$ case
\be
\label{L6}
\lag_6 = \sum_{i=1,94} C_i\,O_i
\ee
and refer to \cite{BCE}, where this was first constructed after the earlier
attempt of \cite{FS},
for the full expressions.
The last four terms are contact terms~\cite{BCE}.

All ultra-violet divergences produced by loop diagrams of order
$p^4$ and $p^6$ cancel in the process of renormalization with the divergences
extracted from the low energy constants $L_i$'s and $C_i$'s.
We use here dimensional regularization and
the standard modified minimal subtraction
$(\overline{MS})$ version used in ChPT. An exhaustive description of the
regularization and renormalization procedure including a description of
the freedom involved can be found in Ref.~\cite{pipi2} and \cite{BCE2}.

The subtraction of divergences is done explicitly by 
\be 
L_i = (c\mu)^{d-4}[\hat{\Gamma}_i \Lambda + L_i^r(\mu)]
\ee
and
\be
C_i = \frac{(c\mu)^{2(d-4)}}{F^2}\left[C_i^r(\mu)-\left(\Gamma_i^{(1)}
+\Gamma_i^{(L)}(\mu)\right) \Lambda-\Gamma_i^{(2)} \Lambda^2\right]
\ee
where $c$ and $\Lambda$ are defined by 
\be 
\ln{c} = -\frac{1}{2}\left[\ln{4\pi}+\Gamma^\prime(1)+1\right]\,,
\ee
\be 
\Lambda = \frac{1}{16\pi^2(d-4)}\,.
\ee
The coefficients $\hat{\Gamma}_i$, $\Gamma_i^{(1)}$ and $\Gamma_i^{(2)}$
are constants while the $\Gamma_i^{(L)}$\,'s are linear combinations of
the $L_i^r(\mu)$\,'s.
Their explicit expressions can all be found in \cite{BCE2} where they have been
calculated in general.

Scalar form factors can be calculated using functional derivatives
w.r.t. the external scalar field $\tilde{s}$ \cite{GL2}.

\section{The scalar form factors: definitions and overview of the calculation}
\label{FF}

\subsection{Definitions}
\label{Definitions}

The scalar form factors for the pions and kaons are defined as follows:
\be
\label{ffdef}
\langle M_2(p)|\bar{q}_i q_j|M_1(q) \rangle  =  F^{M_1 M_2}_{ij}(t) 
\ee
with $t=(p-q)^2$ and $i,j=u,d,s$ being indices in the flavour basis
and $M_1$, $M_2$ a meson state with the indicated momentum.

In the present case of isospin symmetry $m_u=m_d=\hat{m}$, the various pion
scalar form factors obey
\ba
\label{isopions}
F_{S}^\pi(t) &\equiv&  
        2 F^{\pi^0\pi^0}_{uu}(t) =
        2 F^{\pi^0\pi^0}_{dd}(t) =
        2 F^{\pi^+\pi^+}_{uu}(t) =
        2 F^{\pi^+\pi^+}_{dd}(t) =
\nonumber\\
&=&
-2\sqrt{2}F^{\pi^0\pi^+}_{du}(t) =
 2\sqrt{2}F^{\pi^0\pi^-}_{ud}(t)\,,
\nonumber\\
F_{Ss}^\pi &\equiv & F^{\pi^+\pi^+}_{ss} = F^{\pi^0\pi^0}_{ss}\,.
\ea
The kaon currents are related by the following rotations in flavour space
\ba
\label{isokaons}
F_{Su}^K(t) &\equiv& F_{uu}^{K^+K^+(t)} =  F_{dd}^{K^0K^0}(t)\,,
\nonumber\\
F_{Sd}^K(t) &\equiv& F_{dd}^{K^+K^+}(t) =  F_{uu}^{K^0K^0}(t)\,,
\nonumber\\
F_{Ss}^K(t) &\equiv& F_{ss}^{K^+K^+}(t) =  F_{ss}^{K^0K^0}(t)\,,
\nonumber\\
F_{S}^{K\pi}(t) &\equiv& F_{su}^{K^0\pi^-}(t)
 =  \sqrt{2} F_{su}^{K^+\pi^0}(t)\,,
\nonumber\\
F_{Sq}^K(t) & \equiv&F_{Su}^K(t)+F_{Sd}^K(t)\,.
\ea
The other scalar form factors can be related to these using charge conjugation
and time reversal. In (\ref{isopions}) and (\ref{isokaons}) we have
also given the notation we shall use for the form factors in the remainder.

The scalar form factor $F_{S}^{K\pi}(t)$ is proportional to the form factor
$f_0(t)$ used in $K_{\ell3}$ decays.

The scalar form factors can be shown to obey a relation similar to
the Sirlin \cite{Sirlin} relation for the vector form factor
\be
\label{eq:relation}
F_{S}^\pi(t)-2 F_{Ss}^\pi(t)-2 F_{Sd}^K(t)+2 F_{Ss}^K(t)-4 F_{S}^{K\pi}(t)
= \mathcal{O}\left((m_s-\hat m)^2\right)\,.
\ee
The proof of this relation is in App.~\ref{app:relation}.

The values at zero momentum transfer are related to the derivatives of the
masses w.r.t. to quark masses because of the Feynman-Hellman theorem
(see e.g. \cite{GL3})
\ba
F_{S}^\pi(0) =  \frac{\partial}{\partial\hat m}m_\pi^2\,,&&
F_{Ss}^\pi(0) =  \frac{\partial}{\partial m_s}m_\pi^2\,,
\nonumber\\
F_{Su}^K(0) = \frac{\partial}{\partial m_u}m_K^2\,,&&
F_{Ss}^K(0) =  \frac{\partial}{\partial m_s}m_K^2\,,
\nonumber\\
F_{Sd}^K(0) = \frac{\partial}{\partial m_d}m_K^2\,.
\ea

This also shows that we can expect large corrections at $t=0$. The argument
is fairly simple. If
\be
m_K^2 \approx B_0 m_s + \beta (B_0 m_s)^2 + \gamma (B_0 m_s)^3\,,
\ee
then
\be
F_{Ss}^K(0) \approx B_0 + 2 \beta B_0 m_s + 3 \gamma (B_0 m_s)^2\,.
\ee
So we see that in the scalar form factors the relative $p^6$ corrections can
get enhanced by factors of order 3 compared to the masses.

\subsection{Diagrams}
\label{diagrams}

The relevant diagrams for the present case are identical to those involved in
the electromagnetic form factors \cite{BT1} with the electromagnetic current
replaced by the scalar one. We list the diagrams appearing
at each order but refer to \cite{ABT3,BT1} for a deeper discussion of the
checks to perform and the renormalization.

\subsubsection{Leading and next-to-leading orders}
\label{NLO}

\begin{figure}
\begin{center}
\includegraphics[width=0.8\textwidth]{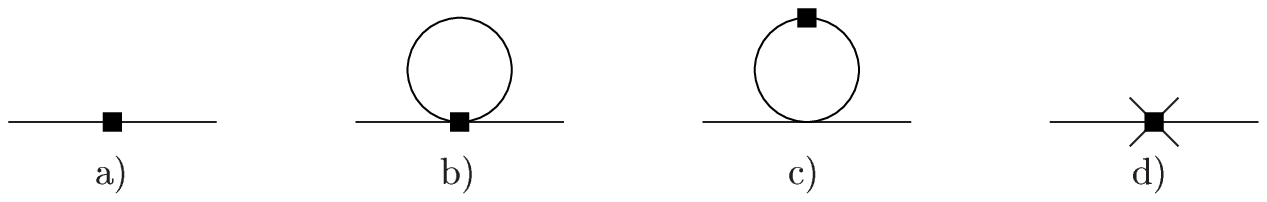}
\end{center}
\caption{Order $p^2$ and $p^4$. \label{p2p4diags}}
\end{figure}

The diagrams are shown in Fig~\ref{p2p4diags}. The black square represents
an external scalar insertion from $\lag_2$ - diagram a) - and the crossed
square a scalar insertion form $\lag_4$ - diagram d). Also included, are the
tadpole contributions with two possible insertions.

The lowest order contributions are
\ba
F^\pi_{S}(t) &=& 2 B_0\,,\quad F^\pi_{Ss}(t) = 0\,,
\nonumber\\
F^K_{Sq}(t) &=&  B_0\,,\quad F^K_{Ss}(t) =  B_0
\nonumber\\
F^{K\pi}_{S}(t) &=&  B_0\,.
\ea

\subsubsection{Next-to-next-to-leading order} 
\label{NNLO}
\begin{figure}
\begin{center}
\includegraphics[width=0.8\textwidth]{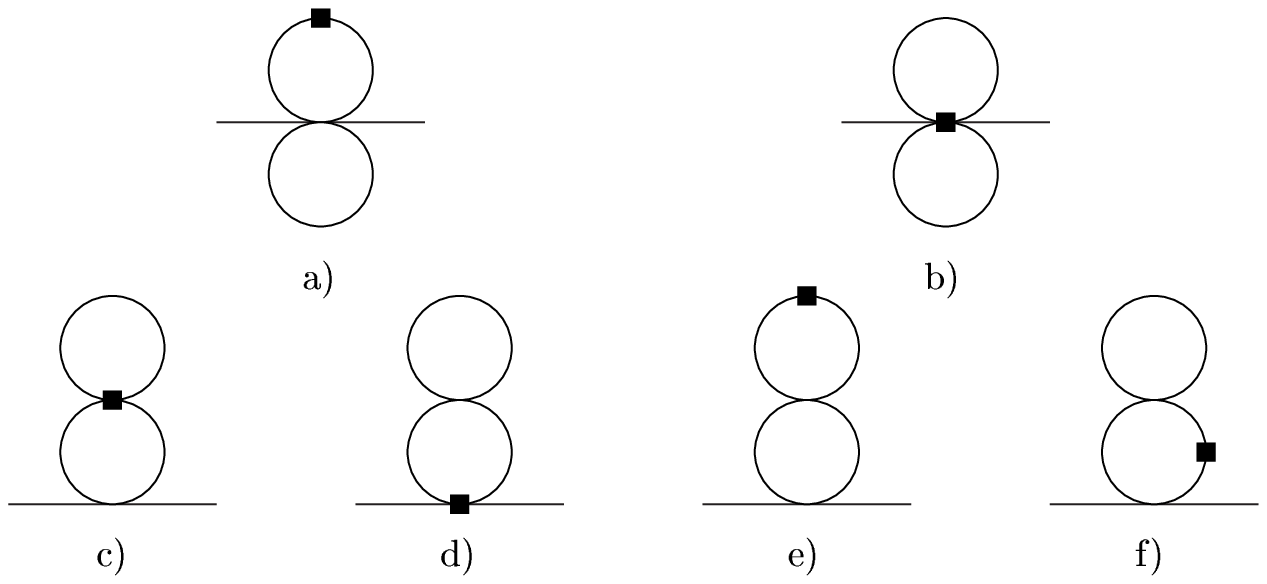}
\end{center}
\caption{The two-loop corrections diagrams. \label{2loopdiags}}
\end{figure}

There are four distinct topologies involved in order $p^6$ diagrams. Firstly,
the two loop diagrams, built exclusively on $O(p^2)$ vertices, are represented
in Fig~\ref{2loopdiags}. There again does a black square represent a scalar
interaction.

\begin{figure}
\begin{center}
\includegraphics[width=0.6\textwidth]{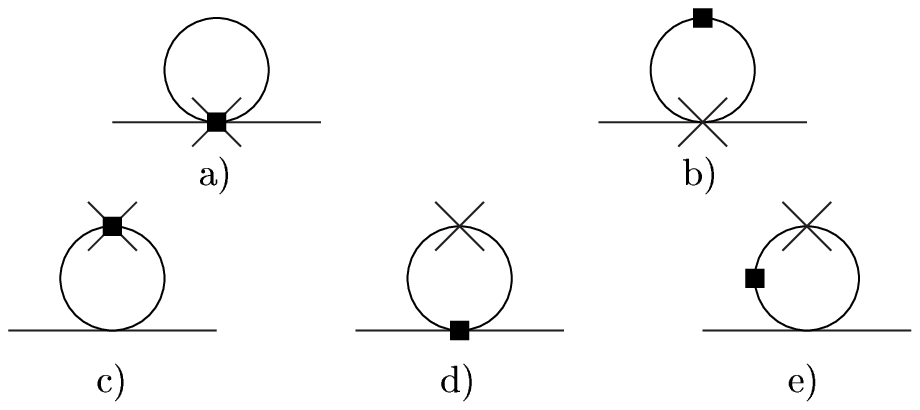}
\end{center}
\caption{One-loop diagrams involving an $O(p^4)$ vertex. \label{1loopdiags}}
\end{figure}

Secondly, one loop diagrams can contribute, with one vertex built on $\lag_4$,
possibly including a scalar insertion. These are represented in
Fig~\ref{1loopdiags}.

\begin{figure}
\begin{center}
\includegraphics[width=0.45\textwidth]{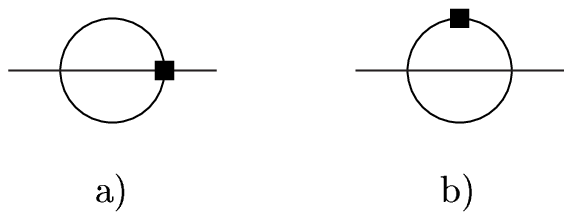}
\end{center}
\caption{Irreducible two-loop diagrams. \label{irreddiags}}
\end{figure}

There are also sunset integrals and irreducible contributions to include
through the non-factorizable diagrams of Fig~\ref{irreddiags}. See \cite{ABT1}
and \cite{BT1} for a treatment of sunset integrals and irreducible two loop
integrals.

\begin{figure}
\begin{center}
\includegraphics[width=0.22\textwidth]{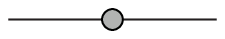}
\end{center}
\caption{The $\lag_6$ contribution. \label{treeL6}}
\end{figure}

Of course, one must finally include the $\lag_6$ tree contribution,
Fig~\ref{treeL6}.

\section{Analytical and First Numerical Results}
\label{results}

\subsection{Analytical expressions - $O(p^6)$ LEC's}
\label{C_contribs}

We present here the dependence of all the scalar form factors on
the LECs of order $p^6$. It can be easily checked that the relation
(\ref{eq:relation}) is satisfied.

\ba
\label{ppuCi}
 \langle \pi | \bar{u}u + \bar{d}d | \pi \rangle &=& \frac{16B_0}{F_\pi^4}
 \Big\{
 m_\pi^2\, m_K^2 \,  (  - 16 C^r_{13} - 8 C^r_{15} + 8 C^r_{16} 
  + 8 C^r_{20} + 72 C^r_{21} + 16 C^r_{32}) 
\nonumber \\ && ~~~
       + m_\pi^2 \,t \,  ( 4 C^r_{12} + 8 C^r_{13} + 2 C^r_{14} + 3 C^r_{15} 
 + 4 C^r_{16} + 2 C^r_{17} + 2 C^r_{34} + 2 C^r_{36} )
\nonumber \\ && ~~~
       + m_\pi^4 \,  (  - 12 C^r_{12} - 16 C^r_{13} - 6 C^r_{14} - 8 C^r_{15} 
 - 14 C^r_{16} - 6 C^r_{17}+18 C^r_{19} 
\nonumber \\ && ~~~~~~
 + 30 C^r_{20} 
  + 30 C^r_{21} 
+ 12 C^r_{31}  + 16 C^r_{32} )
       + 2 m_K^2\, t\,  C^r_{15} 
\nonumber \\ && ~~~
       + m_K^4\,   (  - 8 C^r_{16} + 8 C^r_{20} + 24 C^r_{21} )
       - t^2 \,  ( C^r_{12} + 2 C^r_{13} )\,\Big\}\,.
\ea

\ba 
 \label{ppsCi}
\langle \pi | \bar{s}s | \pi \rangle &=&  \frac{8B_0}{F_\pi^4} \Big\{ 
 m_\pi^2\, m_K^2 \,  (  - 16 C^r_{16} + 16 C^r_{20} + 48 C^r_{21} )
       + m_\pi^2 \,t  \, ( 8 C^r_{13} + 2 C^r_{15} - 4 C^r_{16} + 2 C^r_{36} )
\nonumber \\ && ~~~
       + m_\pi^4 \,  (  - 8 C^r_{13} - 4 C^r_{15} + 8 C^r_{16} + 24 C^r_{21} 
 + 8 C^r_{32} )
       + 8\,m_K^2\, t \,C^r_{16} 
       - 2\, t^2\, C^r_{13} \,\Big\}\,.
\ea
\ba
\label{ppkCi}
 \langle K^+ | \bar{u}u + \bar{d}d | K^+ \rangle &=& \frac{8B_0}{F_\pi^4} 
\Big\{ 
       + m_\pi^2\, m_K^2 \, (  - 8 C^r_{13} - 4 C^r_{15} - 8 C^r_{16} 
         + 16 C^r_{20} + 72 C^r_{21} + 8 C^r_{32} ) 
\nonumber \\ && ~~~
       + m_\pi^2\, t\,   ( 2 C^r_{14} + C^r_{15} + 8 C^r_{16} - 2 C^r_{17} ) 
\nonumber \\ && ~~~
       + m_\pi^4 \,  (  - 2 C^r_{14} - 6 C^r_{16} + 2 C^r_{17} + 6 C^r_{19} 
  + 6 C^r_{20} + 6 C^r_{21} )
\nonumber \\ && ~~~
       + m_K^2\, t \,  ( 4 C^r_{12} + 16 C^r_{13} + 6 C^r_{15} + 4 C^r_{17} 
    + 2 C^r_{34} + 4 C^r_{36} )
\nonumber \\ && ~~~
       + m_K^4 \,  (  - 12 C^r_{12} - 32 C^r_{13} - 4 C^r_{14} -
        16 C^r_{15} - 8 C^r_{16} - 8 C^r_{17} + 12 C^r_{19}
\nonumber \\ && ~~~ 
       + 40 C^r_{20} 
             + 120 C^r_{21} + 12 C^r_{31} + 32 C^r_{32} )
       - t^2 \,  ( C^r_{12} + 4 C^r_{13} ) \,\Big\}
\ea

\ba
\label{ssk} 
\langle K^+ | \bar{s}s | K^+ \rangle &=& \frac{8B_0}{F_\pi^4} \Big\{ 
       + m_\pi^2 \,m_K^2 \,  (  - 8 C^r_{13} + 8 C^r_{14} - 4 C^r_{15} 
+ 16 C^r_{16} - 8 C^r_{17} - 24 C^r_{19} - 8 C^r_{20} 
\nonumber \\ && ~~~
+ 48 C^r_{21} + 8 C^r_{32} )
       + m_\pi^2 \,t \,  (  - 2 C^r_{14} + C^r_{15} - 4 C^r_{16} + 2 C^r_{17} )
\nonumber \\ && ~~~
       + m_\pi^4  \, (  - 2 C^r_{14} - 6 C^r_{16} + 2 C^r_{17} +
       6 C^r_{19} + 6 C^r_{20} + 6 C^r_{21} )
\nonumber \\ && ~~~
       + m_K^2 \,t \,  ( 4 C^r_{12} + 8 C^r_{13} + 4 C^r_{14} 
       + 4 C^r_{15} + 8 C^r_{16} + 2 C^r_{34} + 2  C^r_{36} ) 
\nonumber \\ && ~~~
       + m_K^4 \,  (  - 12 C^r_{12} - 24 C^r_{13} - 12 C^r_{14} - 12 C^r_{15} 
    - 24 C^r_{16} + 36  C^r_{19} + 48 C^r_{20} 
\nonumber \\ && ~~~
    + 72 C^r_{21} + 12 C^r_{31} + 24 C^r_{32} )
       - t^2  \, ( C^r_{12} + 2 C^r_{13} )\Big\}\,.
\ea

\ba
\label{ppk2Ci}
\langle \pi^- | \bar{s}u | K^0 \rangle &=& \frac{8B_0}{F_\pi^4} \Big\{
       + m_\pi^2 \, m_K^2 \,   (  - 2 C^r_{12} - 12 C^r_{13} - 6 C^r_{15} 
+ 8 C^r_{16} - 4 C^r_{17} + 4 C^r_{20}+ 24 C^r_{21} 
\nonumber\\ && ~~~
+ 4 C^r_{31} + 12 C^r_{32} + 2 C^r_{34} )
       + m_\pi^2 \, t \,  ( 2 C^r_{12} + C^r_{15} + 2 C^r_{17} + C^r_{34} )
\nonumber \\ && ~~~
       + m_\pi^4 \,   (  - 5 C^r_{12} - 4 C^r_{13} - 2 C^r_{14} - 2 C^r_{15}
 - 6 C^r_{16} - 2 C^r_{17} + 6 C^r_{19} + 10 C^r_{20}
\nonumber\\ && ~~~
 + 6 C^r_{21} 
+ 4 C^r_{31} + 4 C^r_{32} - C^r_{34} )
       + m_K^2 \, t \, ( 2 C^r_{12} + 2 C^r_{14} + 2 C^r_{15} + C^r_{34} )
 \nonumber \\ && ~~~
       + m_K^4 \, (  - 5 C^r_{12} - 8 C^r_{13} - 4 C^r_{14} - 4 C^r_{15} 
 - 8 C^r_{16} + 12 C^r_{19} + 16 C^r_{20}
 + 24 C^r_{21}
\nonumber \\ && ~~~
 + 4 C^r_{31} + 8 C^r_{32} - C^r_{34} ) 
       - t^2 \, C^r_{12} \Big\}
\ea

\subsection{Loop corrections}
\label{loops}

We present only the results for $F^\pi_{S}$. The expressions for the
others are obtainable on request from the authors. Form factors including
external etas can be calculated as well if required. The order $p^4$
results are in agreement with those of \cite{GL3}. The form given in
App.~\ref{Appresults} is the one that corresponds to the $p^6$ expressions
we use and which can also be found in App.~\ref{Appresults}.

\subsection{First Numerical Results}
\label{plots}

In this part we present some plots of the sizes of the various
corrections for the case with $C_i^r=0$, we have used the $L_i^r$ values
of \cite{ABT4}, fit 10, and the neutral pion, neutral kaon and
physical eta masses as well as $F_\pi = 92.4~$MeV and $\mu = 770~$MeV.
These plots are included here to show the relative sizes of the many
contributions to the scalar form factors we have calculated. These
corrections are in many cases large as expected from the argument
given at the end of Section~\ref{Definitions}.

\begin{figure}
\begin{minipage}{0.49\textwidth}
\includegraphics[height=0.999\textwidth,angle=-90]{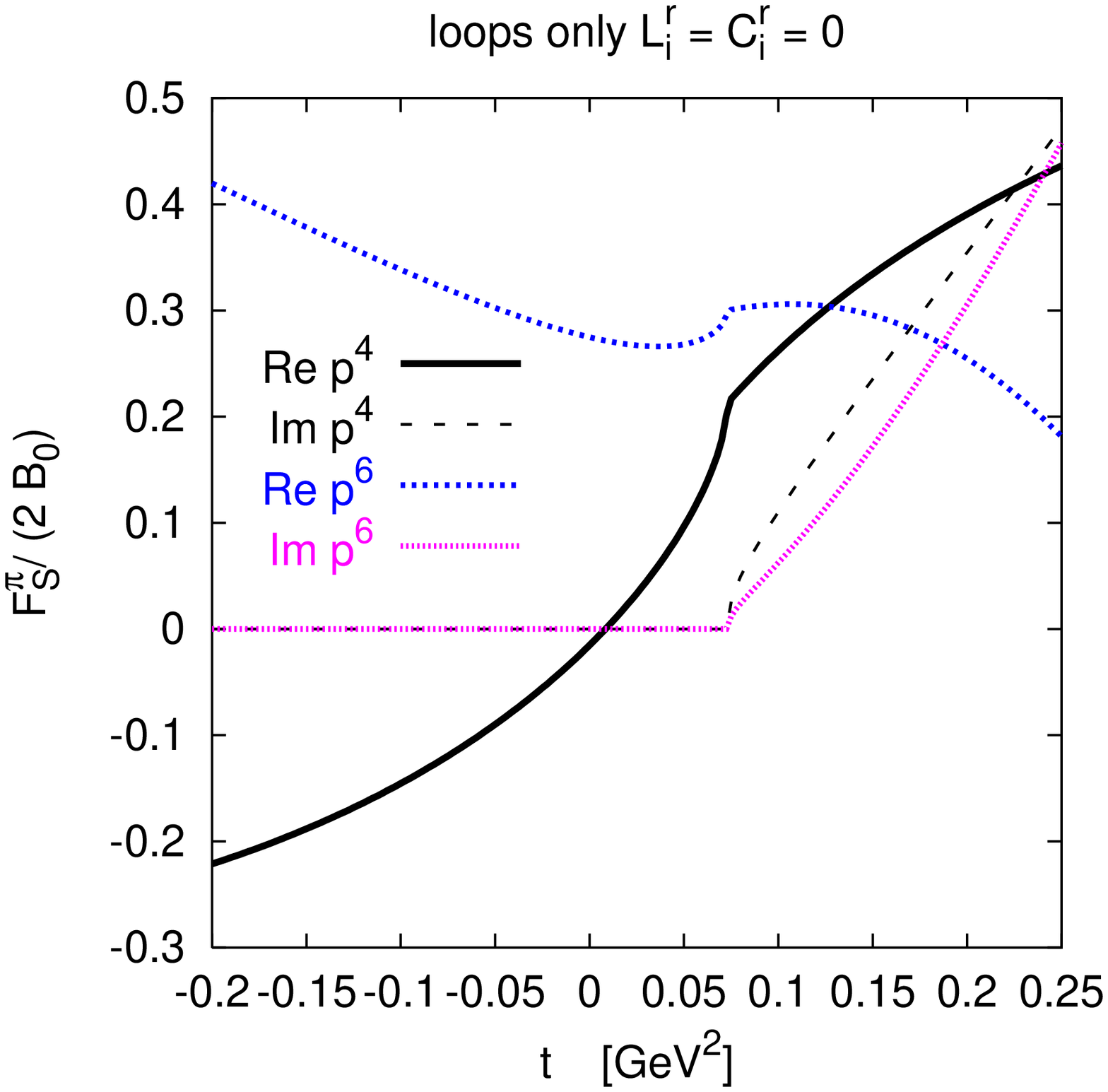}
\centerline{(a)}
\end{minipage}
\hfill
\begin{minipage}{0.49\textwidth}
\includegraphics[height=0.999\textwidth,angle=-90]{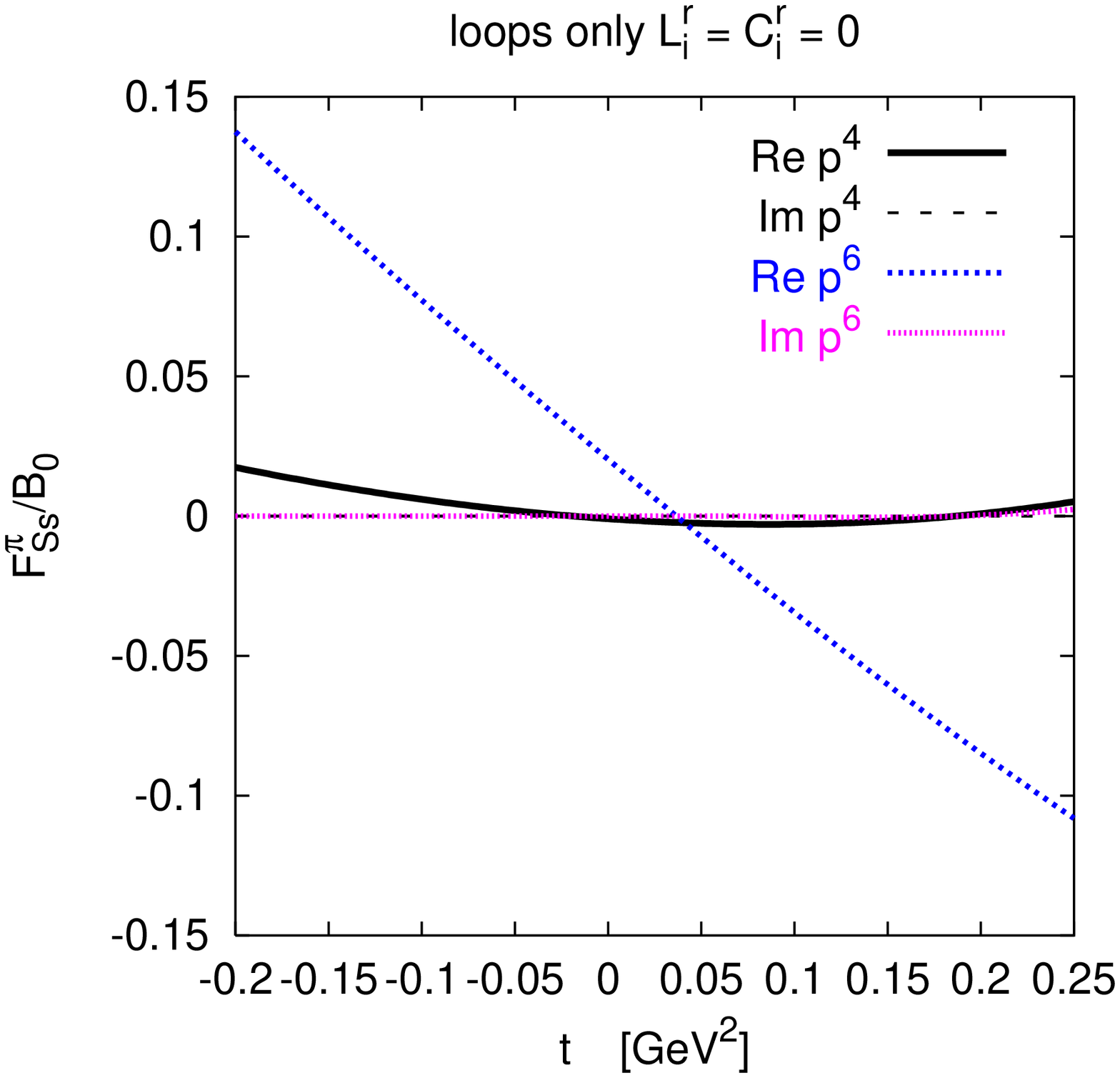}
\centerline{(b)}
\end{minipage}
\caption{The various contributions to (a) $F^\pi_{S}(t)/(2B_0)$
and (b) $F^\pi_{Ss}(t)/B_0$
as a function of $t$ for the case with $L_i^r = C_i^r=0$. Plotted are
the real and imaginary parts of the $p^4$ and the
$p^6$ contributions separately.}\label{figpiqzero}
\end{figure}

In Fig.~\ref{figpiqzero}a we have plotted the loop contributions to the
pion scalar form factor normalized to $2 B_0$. The order $p^2$ is then one.
 The two-loop
contribution in this case is sizable but not enormously so.
The loop contributions to the
pion strange scalar form factor,
$F^\pi_{Ss}(t)/B_0$, are plotted in Fig.~\ref{figpiqzero}b. The
corrections here are moderate. Note that the imaginary part vanishes at
order $p^4$ and remains small at $p^6$. This was expected since it requires
a kaon or eta loop when the LECs are set to zero. 

\begin{figure}
\begin{minipage}{0.49\textwidth}
\includegraphics[height=0.99\textwidth,angle=-90]{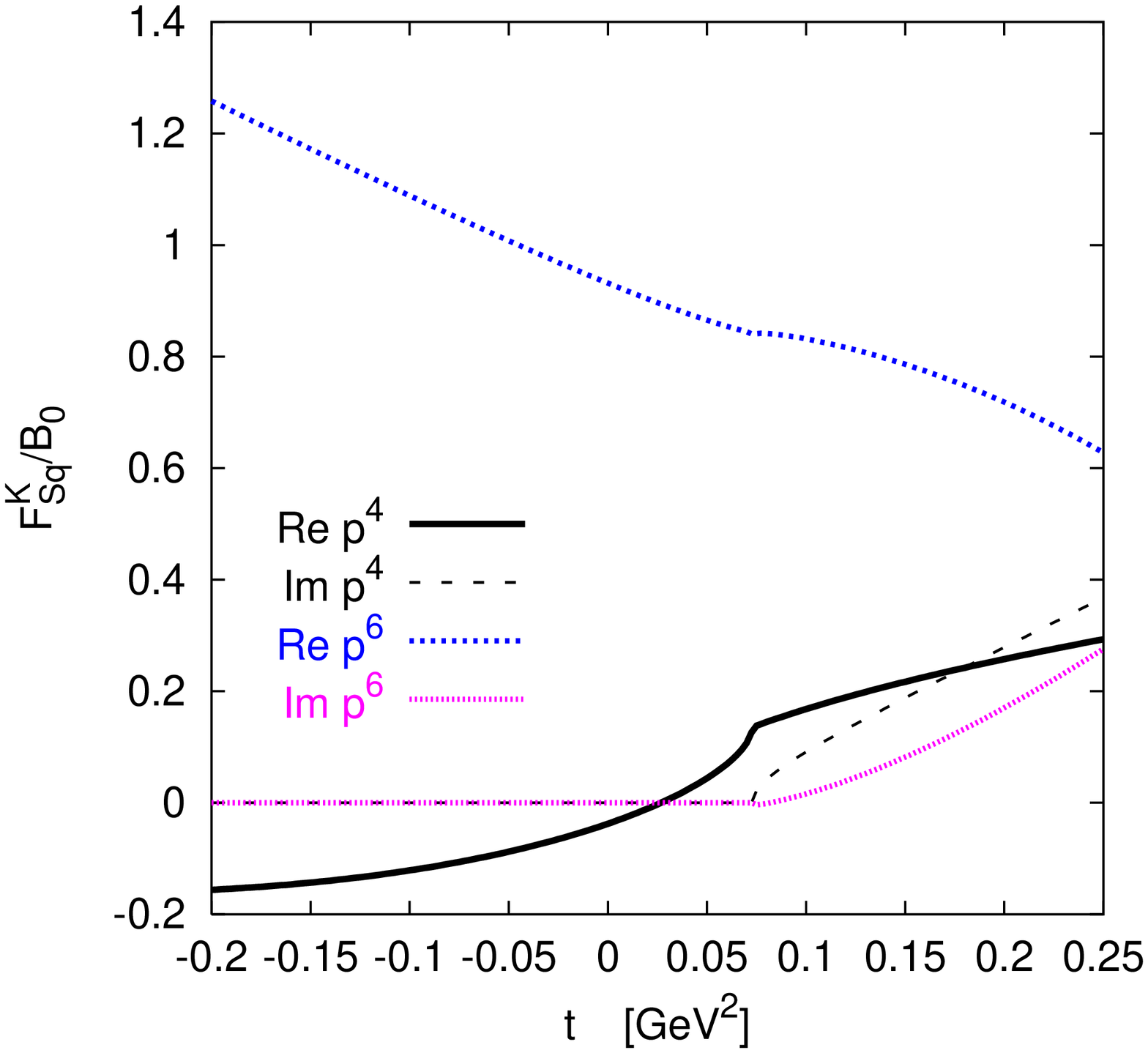}
\centerline{(a)}
\end{minipage}
\begin{minipage}{0.49\textwidth}
\includegraphics[height=0.99\textwidth,angle=-90]{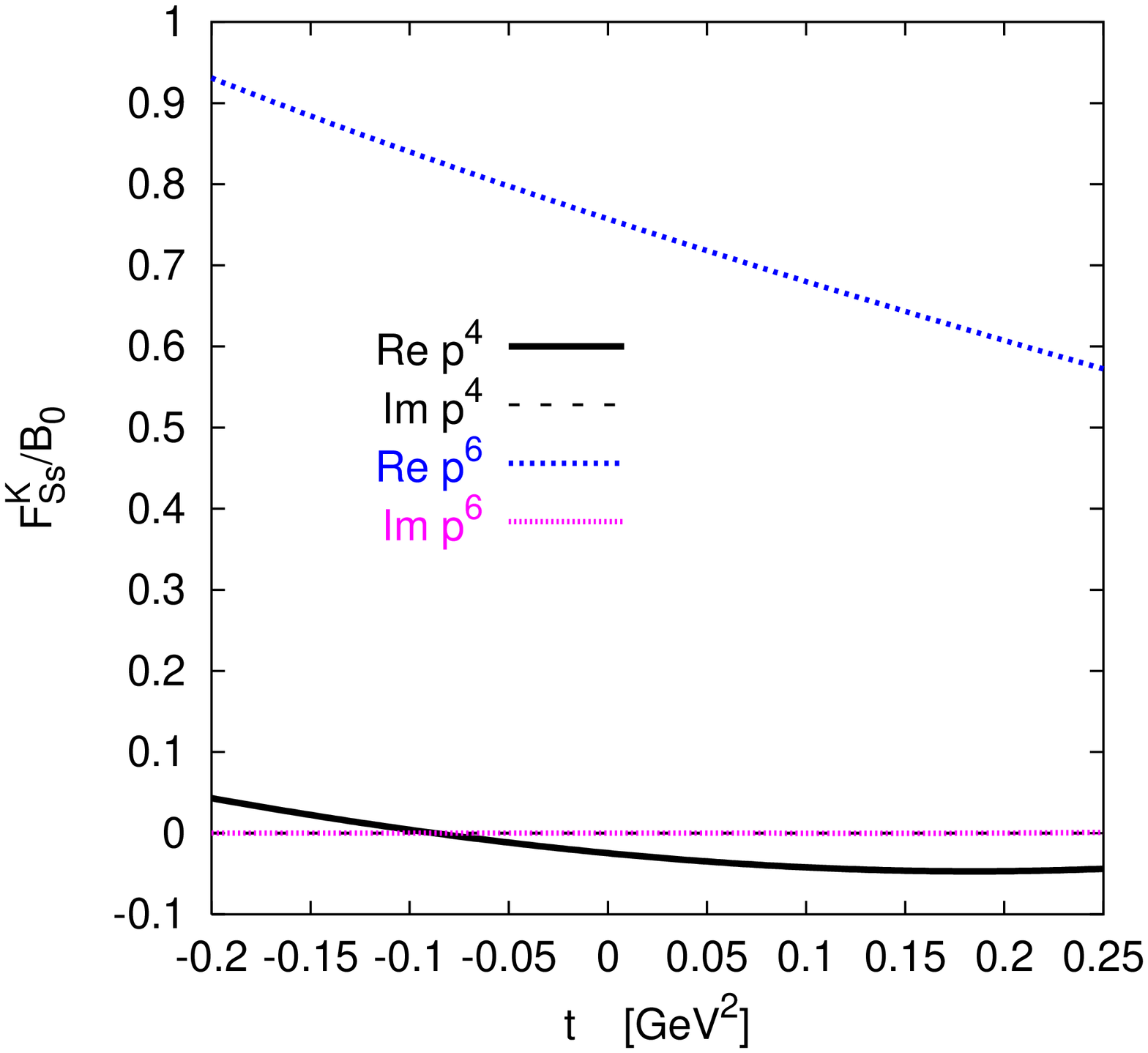}
\centerline{(b)}
\end{minipage}
\caption{The various contributions to (a) $F^K_{Sq}(t)/B_0$
and (b) $F^K_{Ss}(t)/B_0$
as a function of $t$ for the case with $L_i^r = C_i^r=0$. Plotted are
the real and imaginary parts of the $p^4$ and the
$p^6$ contributions separately.}\label{figKqzero}
\end{figure}

In Fig.~\ref{figKqzero}a we have plotted the loop contributions to the
kaon light quark scalar form factor normalized to $B_0$.
There are large two-loop
contributions in this case for the real part while the imaginary part
has only modest corrections. This agrees with the imaginary part
calculated previously in \cite{Meissner1}.

The kaon strange scalar form factor,
$F^K_{Ss}(t)/B_0$, is plotted in Fig.~\ref{figKqzero}b. The
corrections here are also large compared to the order $p^2$ result which is 1.
The imaginary parts vanish at order $p^4$
and remain very small at order $p^6$.

We will do a more extensive analysis in the next section
but here we want to show the effects of the diagrams involving $L_i^r$
separately. We use here the $L_i^r$
of fit 10 of \cite{ABT4},
the $C_i^r(\mu= 770{\rm ~MeV})=0$ and the
other input parameters as given above. We also present the dependence on
$L_4^r$ by changing it from zero to
 $L_4^r(\mu= 770{\rm ~MeV}) = -0.0003$ keeping the others at
the values given by fit 10.

\begin{figure}
\begin{minipage}{0.49\textwidth}
\includegraphics[height=0.99\textwidth,angle=-90]{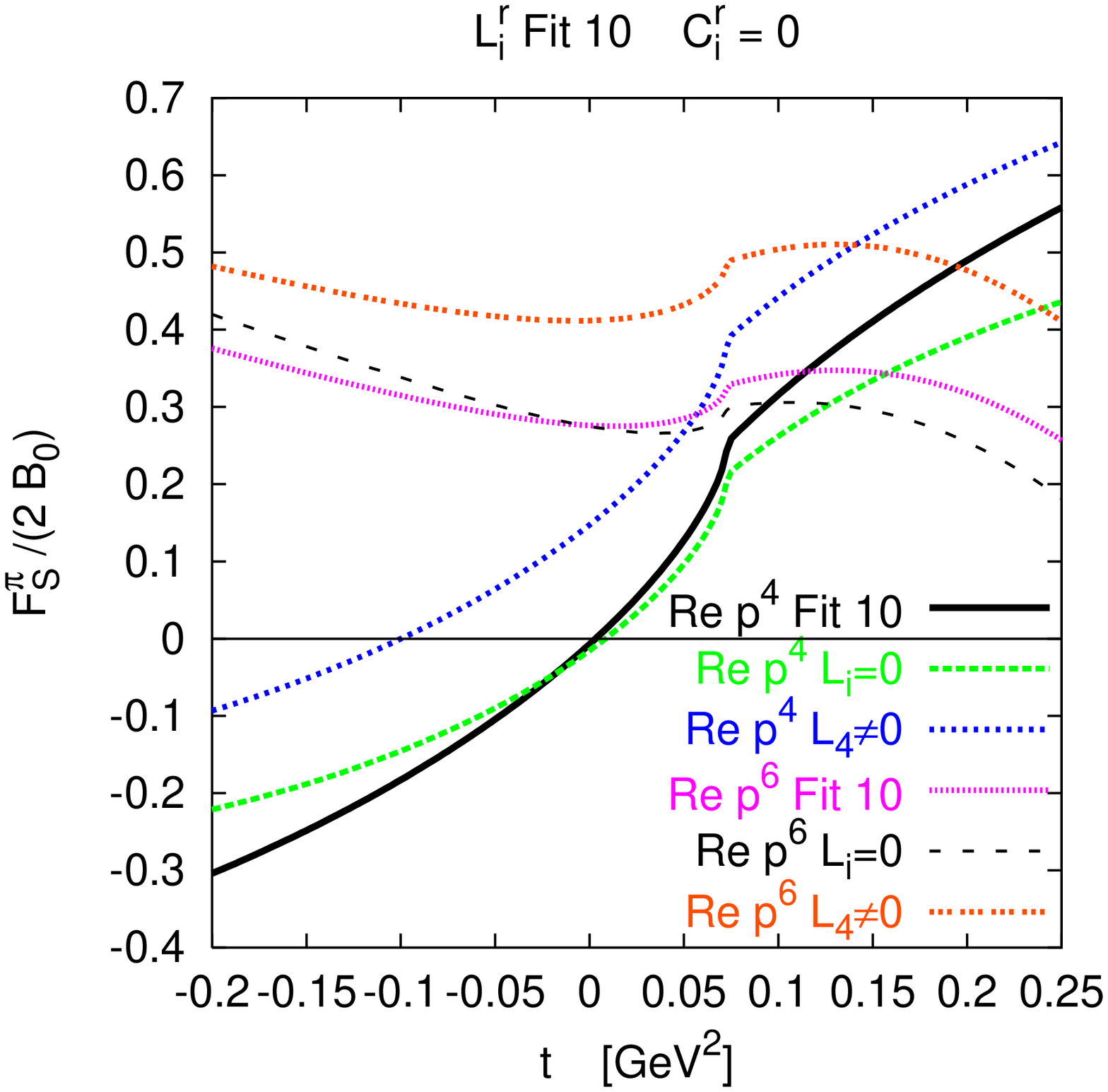}
\centerline{(a)}
\end{minipage}
\begin{minipage}{0.49\textwidth}
\includegraphics[height=0.99\textwidth,angle=-90]{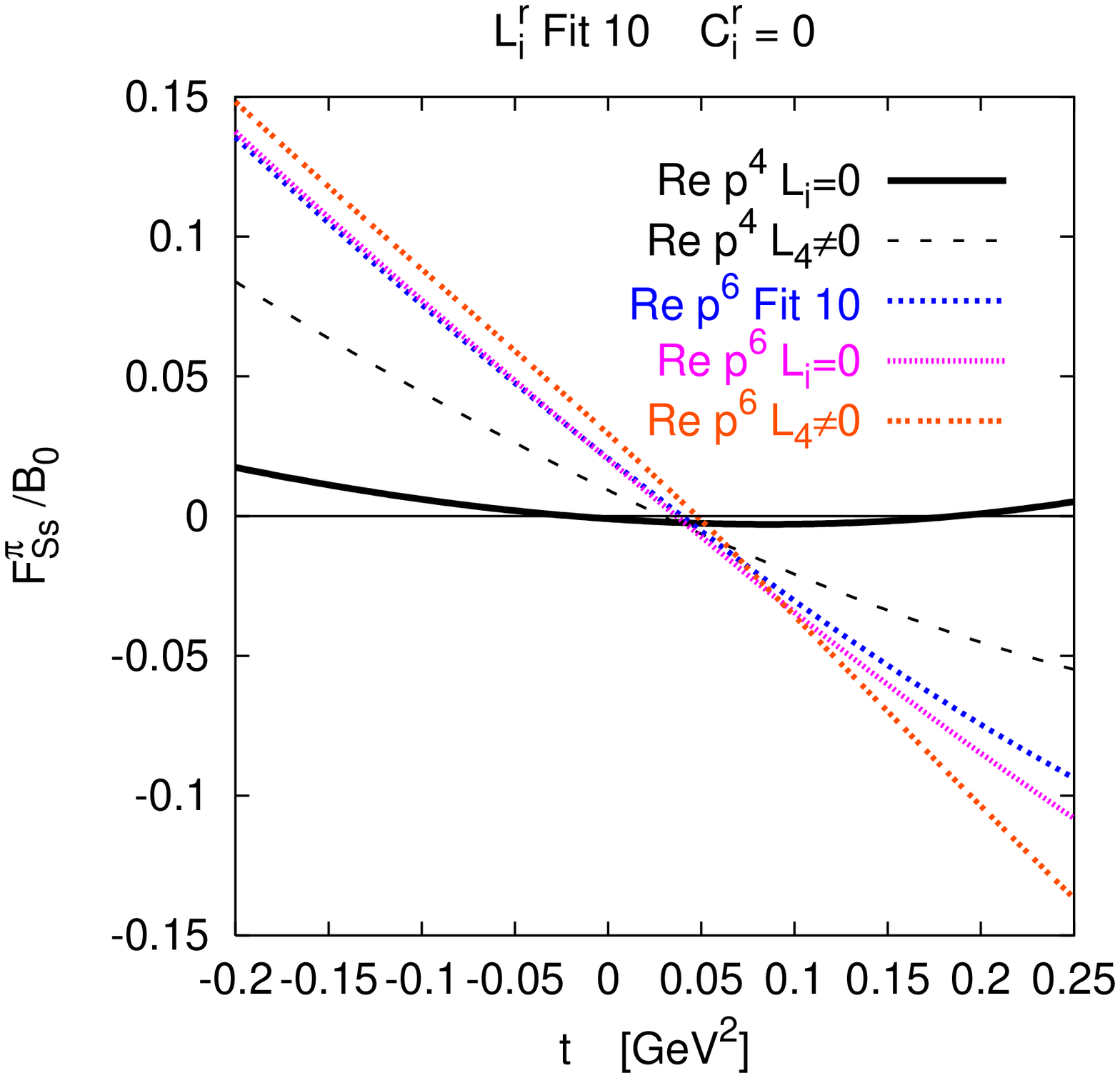}
\centerline{(b)}
\end{minipage}
\caption{The effects of the $L_i^r$ on the various contributions to 
(a) $F^\pi_{S}(t)/(2B_0)$
and (b) $F^\pi_{Ss}(t)/B_0$
as a function of $t$ for the case with $C_i^r=0$. 
The curves are for the real parts
at order $p^4$ and order $p^6$. The curves labelled respectively
Fit 10, $L_4^r\ne0$ and $L_i^r= 0$ are for the
standard values of the $L_i^r$~\cite{ABT4}, the same values but 
$L_4^r = -0.003$ and with all $L_i^r=0$.}
\label{figpiqLi}
\end{figure}

In Fig.~\ref{figpiqLi}a we plotted
the effect of the $L_i^r$ on $F^\pi_{S}(t)/(2B_0)$
compared to the case with the $L_i^r=0$. The curves are for the real parts
at order $p^4$ and order $p^6$. The curves are labelled respectively
Fit 10, $L_i^r=0$ and $L_4^r\ne 0$.
Notice that the effect of such
a small $L_4^r$ is fairly large. 

The pion strange
scalar form factor, $F^\pi_{Ss}(t)/B_0$ is shown in Fig.~\ref{figpiqLi}b.
At order $p^4$ only $L_4^r$ contributes but it has a large effect.
$L_6^r$ contributes as well but is assumed zero. So there is no difference
for this case for Fit 10 and the $L_i=0$.
Notice that
the effect of the $p^6$ corrections is larger than the change due
to this value of $L_4^r$. It is therefore necessary to
include the $p^6$ effects before drawing conclusions on the values
of $L_4^r$ and $L_6^r$.

We can present similar plots for the $L_i^r$ dependence of the kaon scalar
form factors but the conclusions are qualitatively the same as for the pion 
form factors.

\section{Numerical Analysis}
\label{fullnumerics}

\subsection{Dispersive Inputs}
\label{experiments}

There is no direct experimental measurement of the scalar form factors
but information can still be extracted from elastic scattering experiments. 
We recall here the principal results of dispersive analyses and refer to
\cite{DGL} and \cite{Moussallam1} for a more thorough treatment.
An analysis within a unitarized model approach can be found in
\cite{Meissner2}.

\subsubsection{The Muskhelishvili-Omn\`es (MO) problem}
\label{MO1}

We consider first the one channel case, {\em i.e.} where only pion
interactions are considered. The problem of finding the functions
satisfying the pion form factor's analytical properties is referred
to as the Muskhelishvili-Omn\`es problem. Provided some convergence
conditions on the form factor and the phase shift $\delta_\pi(s)$
are satisfied, the solution is of the form:
\be
\label{omnessol}
F_\pi(s)=P(s)\,\exp{\left[\frac{s}{\pi}\int_{s_0}^\infty ds^\prime 
\frac{\delta_\pi(s^\prime)}{s^\prime\,(s^\prime-s)}\right]}.
\ee
Watson's theorem then relates the phase shift to the scattering phases
$\delta_I^J$, here $\delta_\pi=\delta_0^0$. This can be extended to two
or more channels. We restrict here to the two channel case, $\pi\pi$
and $KK$ in the isospin and angular momentum zero case, $I=0$ $S$-wave.
The MO problem becomes a system of coupled equations of the two independent
contributions
\be\label{F1F2}
F_1(s)=\langle 0|\hat{X}_0|\pi\pi\rangle
\hspace{1cm}\mbox{ and }\hspace{1cm}
F_2(s)=\frac{2}{\sqrt{3}}\langle 0|\hat{X}_0|K\bar{K}\rangle\,,
\ee
where $\hat{X}_0$ is an isospin zero operator, 
{\em e.g.} $(\bar{u}u+\bar{d}d)\mbox{ or }\bar{s}s$. The interesting feature
of the MO solution is the possibility to write the general solution
using two independent sets of solutions 
$\{F_1^{(i)},F_2^{(i)}\}_{i=1,2}$ fulfilling the initial conditions
\be
\label{F1F2orig}
F_i^{(j)}(0)=\delta_{ij},
\ee
and this {\em independently} of the exact form of the operator
$\hat{X}_0$. The solution is then built on the value of the form
factors at the origin:
\be
\label{F1F2sol}
\left\{ \begin{array}{ccc}
\dsp F_1(s)\quad= &\dsp  F_1(0)\,F_1^{(1)}(s) &+\quad F_2(0)\,F_1^{(2)}(s)
\\[1mm]
\dsp F_2(s)\quad= &\dsp  F_1(0)\,F_2^{(1)}(s) &+\quad F_2(0)\,F_2^{(2)}(s)
\end{array}\right.
\ee

\subsubsection{The solutions}
\label{MO2}

The method and coding of the solutions were borrowed from Moussallam 
\cite{Moussallam1,ABM},
where the solutions are obtained by solving the linear system of
equations -- equivalent to (\ref{omnessol}) for the two channel case --
given by the discretized form of
\be
\label{analycity}
\mbox{Re}\,F_i(s)=PV\int_{4m_\pi^2}^\infty ds^\prime 
\frac{\mbox{Im}\,F_i(s)}{s^\prime-s}
\ee
and 
\be\label{2channel}
\mbox{Im}\,F_i(s)=2i\sum_{j=1,2}\sigma_j(s)T_{ij}(s)F_j(s)^*
\ee
where
\be\label{sigma}
\sigma_j(s)=\sqrt{1-\frac{4m_j^2}{s}}\,\theta(s-4m_j^2)\,,\quad m_j=m_\pi,m_K
\ee
and $T_{ij}$ are the T-matrix elements of the 
needed $\pi\pi$, $KK$ scattering channels. 
Three T-matrix models of scattering are used,
from Au {\em et al.} \cite{Au},
Kami\'nski {\em et al.} \cite{KKL}
and Ananthanaryan et al. \cite{ABM},
and we refer again to \cite{Moussallam1,ABM} 
for details and differences.
\begin{table}
\begin{center}
\begin{tabular}{llllllll}
\hline
\rule{0cm}{3ex}
 & $F_1(0)$ & $F_1^\prime(0)$ & $1/2\,F_1^{\prime\prime}(0)$ & $F_2(0)$ 
 & $F_2^\prime(0)$ & $1/2\,F_2^{\prime\prime}(0)$\\[0.3ex]
\hline
\rule{0cm}{3ex}Au$^{(1)}$ & 1. & 2.33 & 9.94 & 0. & 0.92 & 5.31 \\
Au$^{(2)}$ & 0. & 0.31 & 1.03 & 1. & 0.78 & 0.98 \\
\rule{0cm}{3ex}lm$^{(1)}$ &  1. & 2.43 & 10.28 & 0. & 0.75 & 3.49 \\
lm$^{(2)}$ &  0. & 0.27 & 0.93 & 1. & 0.84 & 1.00\\
\rule{0cm}{3ex}Abm$^{(1)}$ & 1. & 2.45 & 10.61 & 0. & 1.03 & 5.36\\
Abm$^{(2)}$ & 0. & 0.21 & 0.72 & 1. & 0.81  & 0.98\\[0.3ex]
\hline
\end{tabular}
\end{center}
\caption{Results of the numerical computation of the Omn\`es solutions for
the different $T_{ij}$ models. 
The superscript
refers to the respective inequivalent sets of solutions.
The first two solutions are from Ref.~\cite{Moussallam1}, the last from
\cite{ABM}.
\label{omnesres}}
\end{table}
The results of our calculation for the different derivatives of the canonical
solutions using the programs of \cite{Moussallam1,ABM}
at the origin are presented in Table~\ref{omnesres}.
We have only used the last solution, since they use a better physical model
for the  $\pi\pi$ to $KK$ amplitude and newer $\pi\pi$ phases.

Similar work exists for the $K\pi$ form factor, $F^{K\pi}_S(t)$, in
\cite{JOP}. We have not obtained their results in a form we can
use in the same way.

\subsection{Other experimental inputs and Values of the $L_i^r$, $i\ne4,6$}
\label{otherinput}

The other experimental inputs are the same as those used in \cite{ABT4}
corresponding to fit 10 there. This differed from the main fits reported there
by using the then preliminary data of \cite{E865}.

We would like to study the dependence on $L_4^r$ and $L_6^r$ as well.
A study was already performed in \cite{ABT3} where we varied the input
assumptions of $L_4^r$ and $L_6^r$ and then refitted the other $L_i^r$.
We have redone this now with the data of \cite{E865} included, so we use all
the inputs described above with as input values for
$L_4^r(m_\rho)$ and $L_6^r(m_\rho)$ values on the grid
\be
\label{grid}
10^4\,L_4^r = -4,\ldots,6
\quad\mbox{and}\quad
10^4\,L_6^r = -3,\ldots,6\,.
\ee
In all these cases we have used the resonance saturation contributions to
the $K_{\ell4}$ constants of order $p^6$ as derived in \cite{ABT3}
and those for
the decay constants and the masses as derived in \cite{ABT1} but we have
chosen the value
of $d_m$ as defined in \cite{ABT1}
to be zero. This is also the choice made in \cite{ABT3}.
The reason for this is that the estimate of the
parameter of \cite{ABT1} was very uncertain and the naive value obtained there
led to $p^6$ contributions to the masses which were enormous.

Fits of a quality similar to the one with $L_4^r=L_6^r=0$ could be
obtained for most points on the grid (\ref{grid}).
All those satisfying
\be
L_6^r \lesssim 0.6 L_4^r + 0.0004\,.
\ee
had a value of $\chi^2$ below 1. The precise definition of the $\chi^2$
can be found in \cite{ABT3,ABT4}. The data used up till now do not allow
for a determination of $L_4^r$ and $L_6^r$. But note that while these
cannot be determined, a given assumption on their value leads to correlated
values of all the other $L_i^r$ in order to obtain a decent fit to the data.
This effect is taken into account in the next subsections.

In the numerical results quoted below
we have used the neutral kaon and pion masses and a subtraction scale
$\mu = 770~$MeV.

\subsection{Variation with $L_4^r$ and $L_6^r$ of the scalar form factors
at $t=0$.}
\label{L4L6}

It should be kept in mind that the values of the other $L_i^r$ used
change as well with the values of $L_4^r$ and $L_6^r$ in accordance
with the fit to the other experimental values as described in 
Section~\ref{otherinput}. The $C_i^r(m_\rho)$ contributing to the scalar
form factors are set to zero here.

\begin{figure}
\begin{minipage}{0.47\textwidth}
\begin{center}
\includegraphics[height=0.99\textwidth,angle=-90]{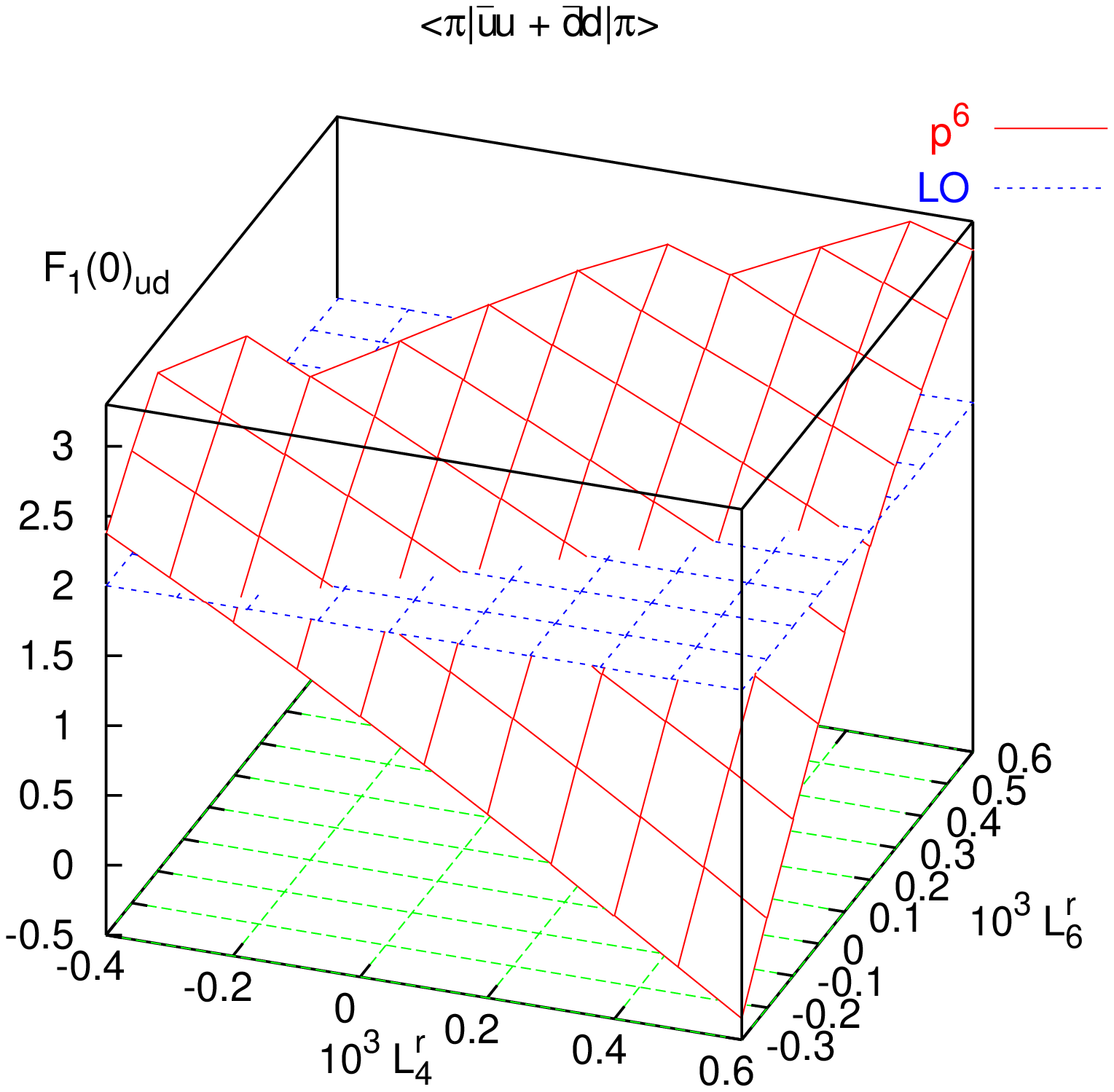}
\\(a)
\end{center}
\end{minipage}
\begin{minipage}{0.47\textwidth}
\begin{center}
\includegraphics[height=0.99\textwidth,angle=-90]{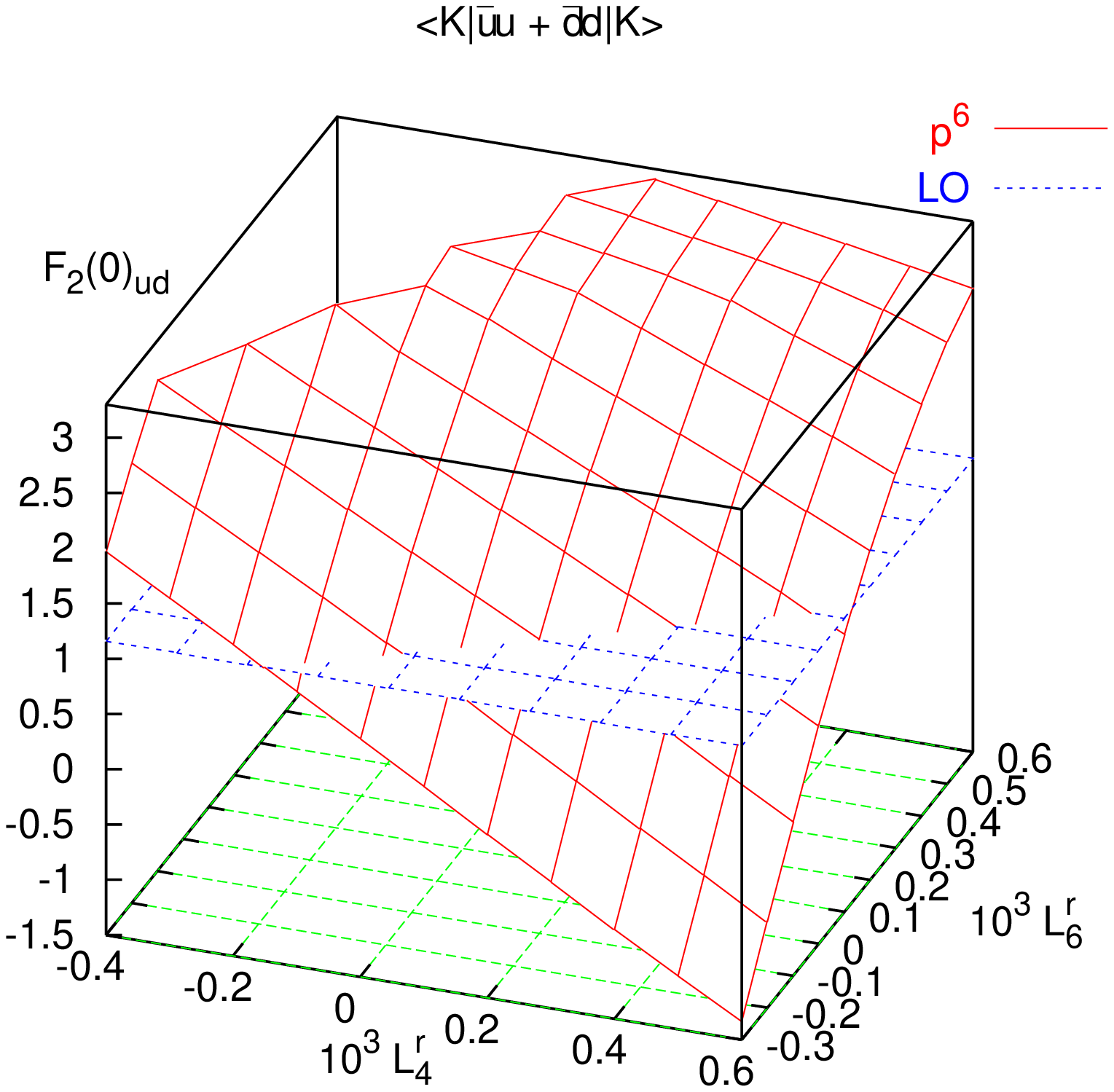}
\\(b)
\end{center}
\end{minipage}
\caption{(a) The result for $F_1(0)_{ud}=F^\pi_{S}(0)/B_0$ as a function of
$L_4^r$ and $L_6^r$. Also shown is the lowest order value $2$.
(b) The result for $F_2(0)_{ud}=2/\sqrt{3}\,F^K_{Sq}(0)/B_0$ as a
function of $L_4^r$ and $L_6^r$. Also shown is the lowest order value
 $2/\sqrt{3}$.}
\label{figF10ud}
\end{figure}

We first show the results for $F^\pi_{S}(0)/B_0$ in Fig.~\ref{figF10ud}a.
We also show the lowest order expectation of 2 for comparison. Only the
points with good fits are shown.

We show similarly in Fig.~\ref{figF10ud}b the value of
$F_2(0)_{ud} =2/\sqrt{3}\,F^K_{Sq}/B_0$ together with the lowest order
value of
$2/\sqrt{3}$.
Notice that in both cases the corrections are large for most values
of $L_4^r$ and $L_6^r$ but reasonable in the region of
the axis $L_6^r \simeq L_4^r - 0.0003$. Requiring the corrections to
be of a reasonable size thus leads to constraints on $L_4^r$ and $L_6^r$.

\begin{figure}
\begin{minipage}{0.475\textwidth}
\begin{center}
\includegraphics[height=0.99\textwidth,angle=-90]{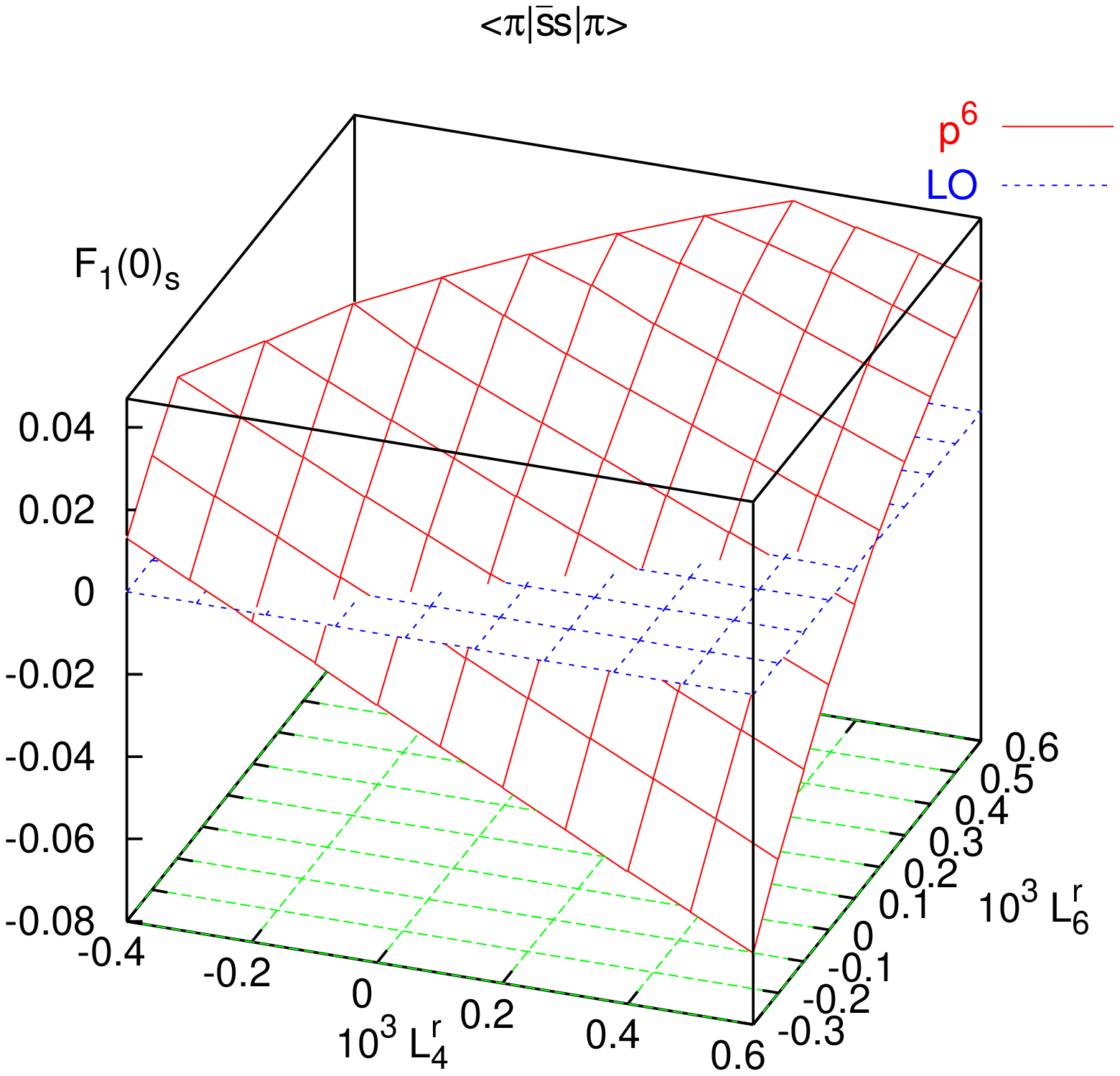}
\\(a)
\end{center}
\end{minipage}
\begin{minipage}{0.475\textwidth}
\begin{center}
\includegraphics[height=0.99\textwidth,angle=-90]{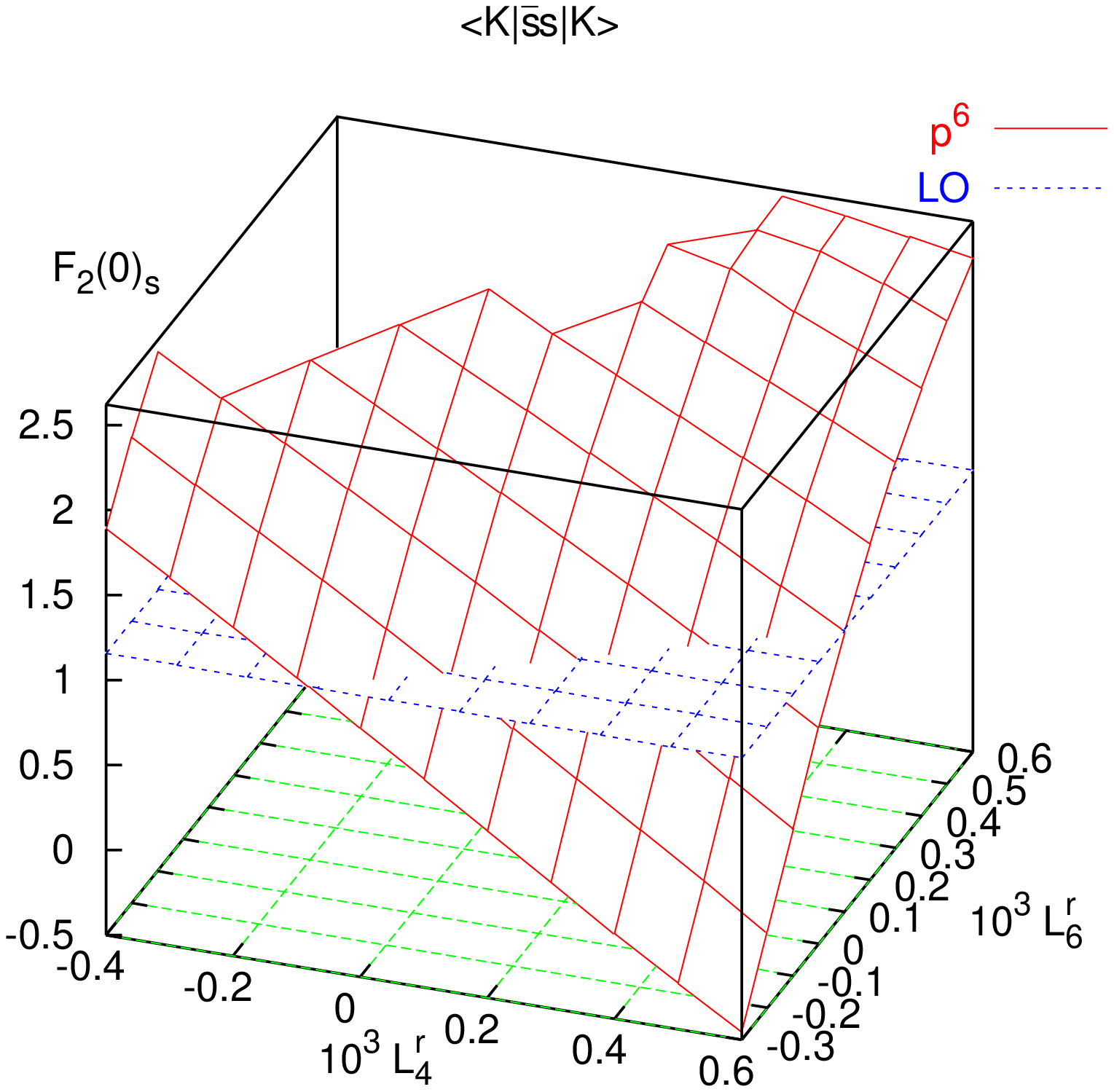}
\\(b)
\end{center}
\end{minipage}
\caption{ (a) The result for $F_1(0)_{s} = F^\pi_{Ss}(0)/B_0$ as a
function of $L_4^r$ and $L_6^r$. Also shown is the lowest order value
 $0$. Notice the strange quark content of the pion remains small
over the entire studied area.
(b) The result for $F_2(0)_{s} = 2/\sqrt{3}\,F^K_{Ss}(0)/B_0$ as a
function of $L_4^r$ and $L_6^r$. Also shown is the lowest order value
 $2/\sqrt{3}$.}
\label{figF10s}
\end{figure}

We have also studied the $\overline{s}s$ current. The results for the pion
case are shown in Fig.~\ref{figF10s}a. 
The result for $F_2(0)_{s} = 2/\sqrt{3}\,F^K_{Ss}(0)/B_0$ is shown in
Fig.~\ref{figF10s}b. Notice that the strange quark content
in the pion remains small for the entire studied region.

At this point we cannot determine the values of $L_4^r$ and $L_6^r$
but there is also a tendency in those two cases for the corrections compared 
to the lowest order to be small for positive $L_4^r$ and a correlated 
value for $L_6^r$ along $L_6^r \simeq L_4^r - 0.0003$. 

\subsection{Slopes and the value of $L_4^r$ and $L_6^r$}
\label{slopes}

We can use our results to get at the form factors away from zero
in two different ways. First we can of course simply use our full
ChPT calculation to calculate them to order $p^6$ using the input values of
$L_4^r$ and $L_6^r$ and the resulting fitted values of the other $L_i^r$.

Second, from
Sect.~\ref{experiments} we can calculate the form factors at $t\ne 0$ given
the values at zero, this uses then all the dispersive constraints.

We then require that both methods  give the same results and obtain
in this way constraints on the values of $L_4^r$ and $L_6^r$. The main
effect of this comparison can already be studied by looking at the
radii only since for the curvature we expect a larger effect from the
$C_i^r$. We still assume here that the $C_i^r(m_\rho)=0$ for those
contributing to the radii.

\begin{figure}
\begin{minipage}{0.475\textwidth}
\begin{center}
\includegraphics[height=0.99\textwidth,angle=-90]{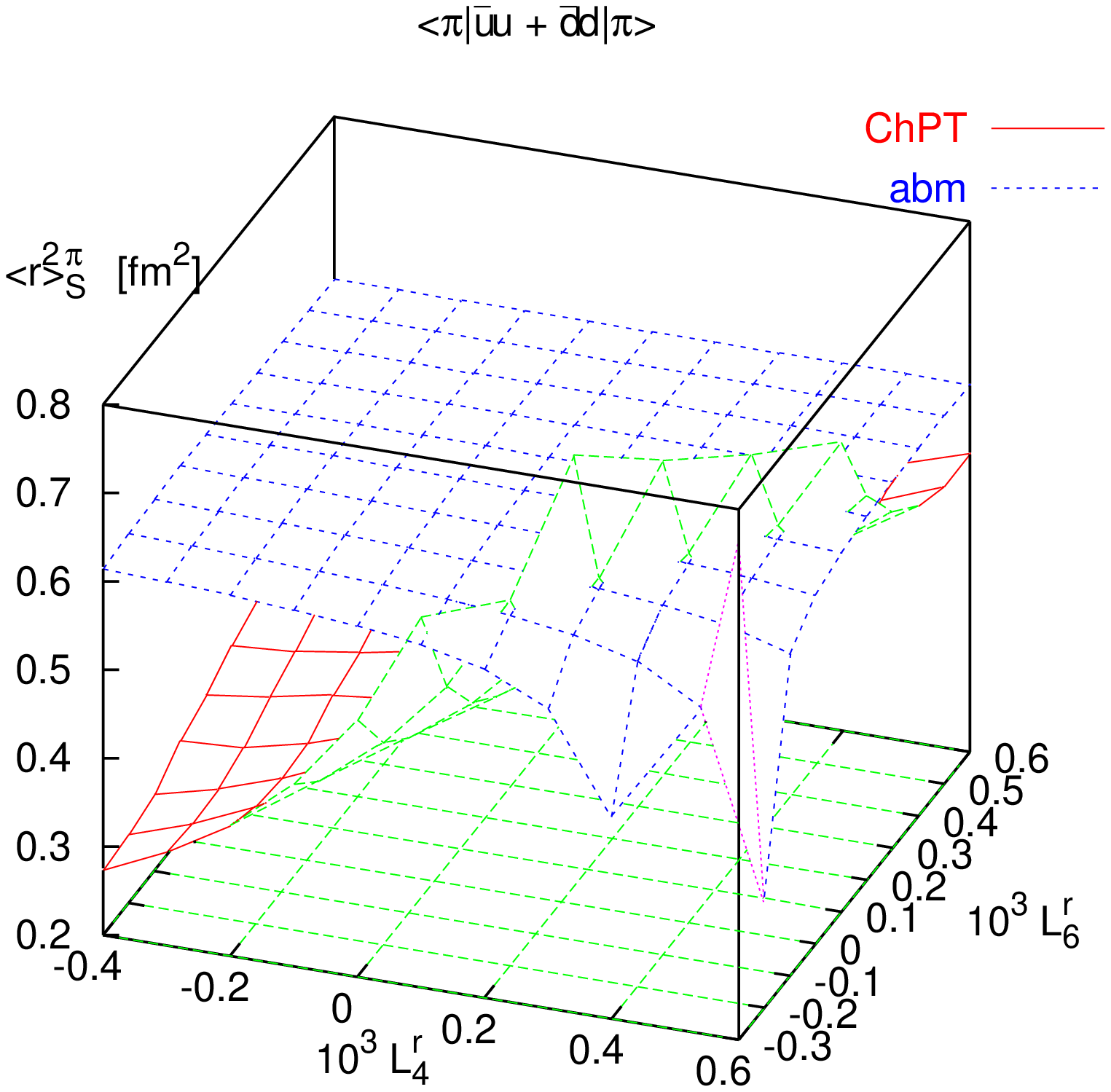}
\\(a)
\end{center}
\end{minipage}
\begin{minipage}{0.475\textwidth}
\begin{center}
\includegraphics[height=0.99\textwidth,angle=-90]{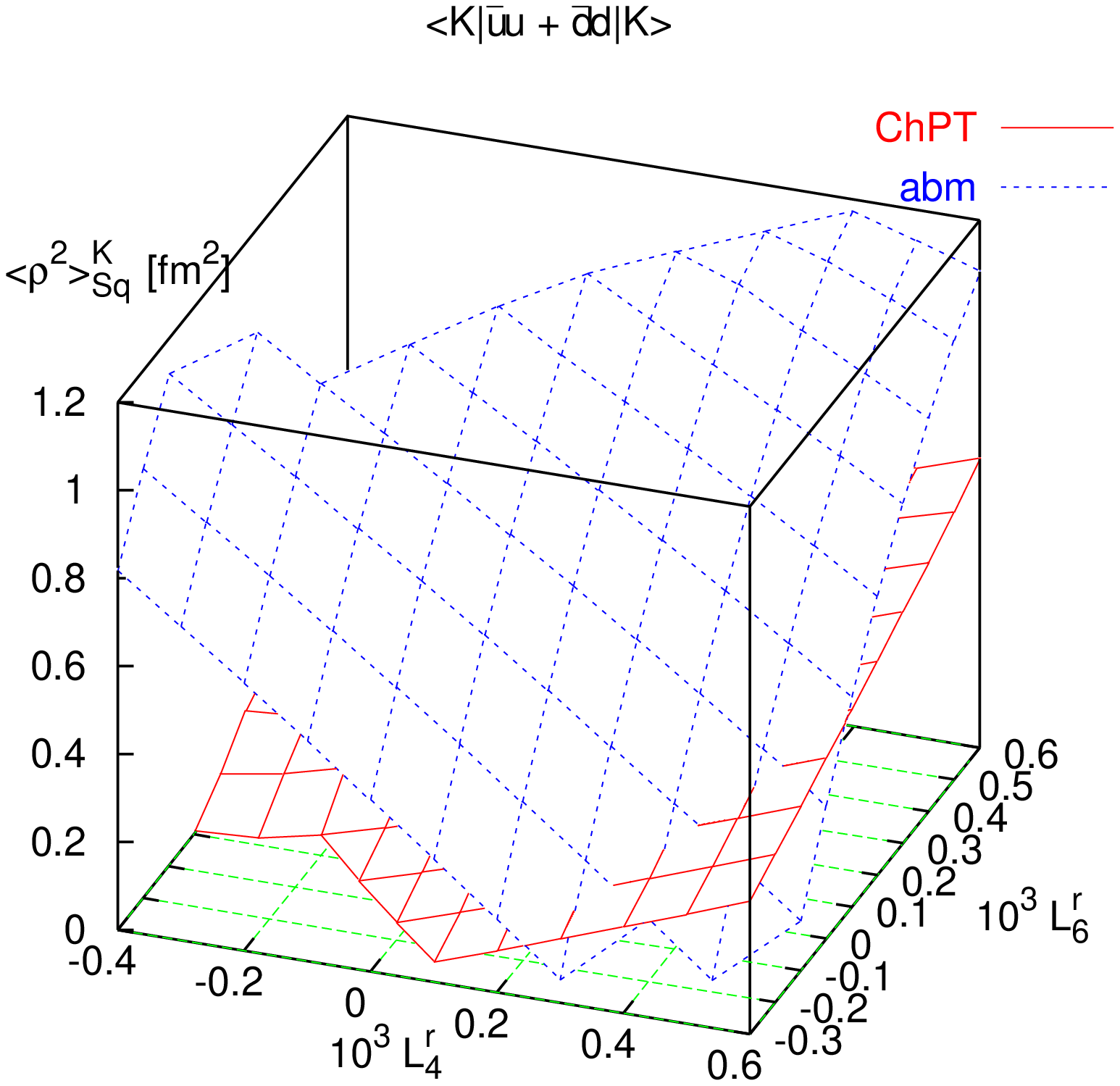}
\\(b)
\end{center}
\end{minipage}
\caption{(a) The result for the scalar radius $<r^2>_\pi^S$
as a function of  $L_4^r$ and $L_6^r$, with the dispersive estimate (abm)
and the ChPT calculation.
(b) The result for the kaon light quark scalar radius 
$<\rho^2>_{Kq}^S$
as a function of  $L_4^r$ and $L_6^r$, with the dispersive estimate (abm)
and the ChPT calculation.}
\label{figrpi2}
\end{figure}

The results for the pion scalar radius,
\be
<r^2>^\pi_S  =  \frac{6}{ F^\pi_{S}(0)}\d{t} F^\pi_{S}(t)|_{t=0}\,,
\ee
are shown in Fig.~\ref{figrpi2}a. We have normalized here to the value at zero.
That leads to singular predictions when $F^{\pi}_{S}(0)$ vanishes,
but away from that region the dispersive prediction
is quite stable and is
entirely predicted by the values of the $\pi\pi$ phase shifts. It is
fully dominated by the first independent solution of the MO equations.
The ChPT
prediction is quite dependent on $L_4^r$ and $L_6^r$ The two agree
for positive values of $L_4^r$ and a correlated value of $L_6^r$.

The results for the kaon light quark scalar radius,
\be
<r^2>_{Sq}^K  =  \frac{6}{ F^K_{Sq}(0)}\d{t} F^K_{Sq}(t)|_{t=0}\,,
\ee
are not as easily shown, as the calculated value of  
$F^K_{Sq}(0)$ runs through zero in the relevant region, producing
strong effects in both the dispersive and ChPT results. We plot
instead the scalar radius normalized to the lowest order result
\be
<\rho^2>_{Sq}^K  =  \frac{6}{B_0}\d{t} F^K_{Sq}(t)|_{t=0}\,.
\ee
This is shown in Fig.~\ref{figrpi2}b. The dispersive prediction is
less stable here. The relative strength of the two canonical MO solutions
varies much more than in the pion case. This has a strong effect
since both solutions have a very different radius, as
can be seen by comparing the first 
derivatives columns in Table~\ref{omnesres}.

Notice that the agreement between the chiral and dispersive
radii tend to be in the same region of $L_4^r$ and $L_6^r$
where the corrections to the lowest order
results for the form factors at $t=0$ are fairly small, as discussed
in the previous subsection, taking into account that the results are
less reliable in the kaon case.

\begin{figure}
\begin{minipage}{0.475\textwidth}
\begin{center}
\includegraphics[height=0.99\textwidth,angle=-90]{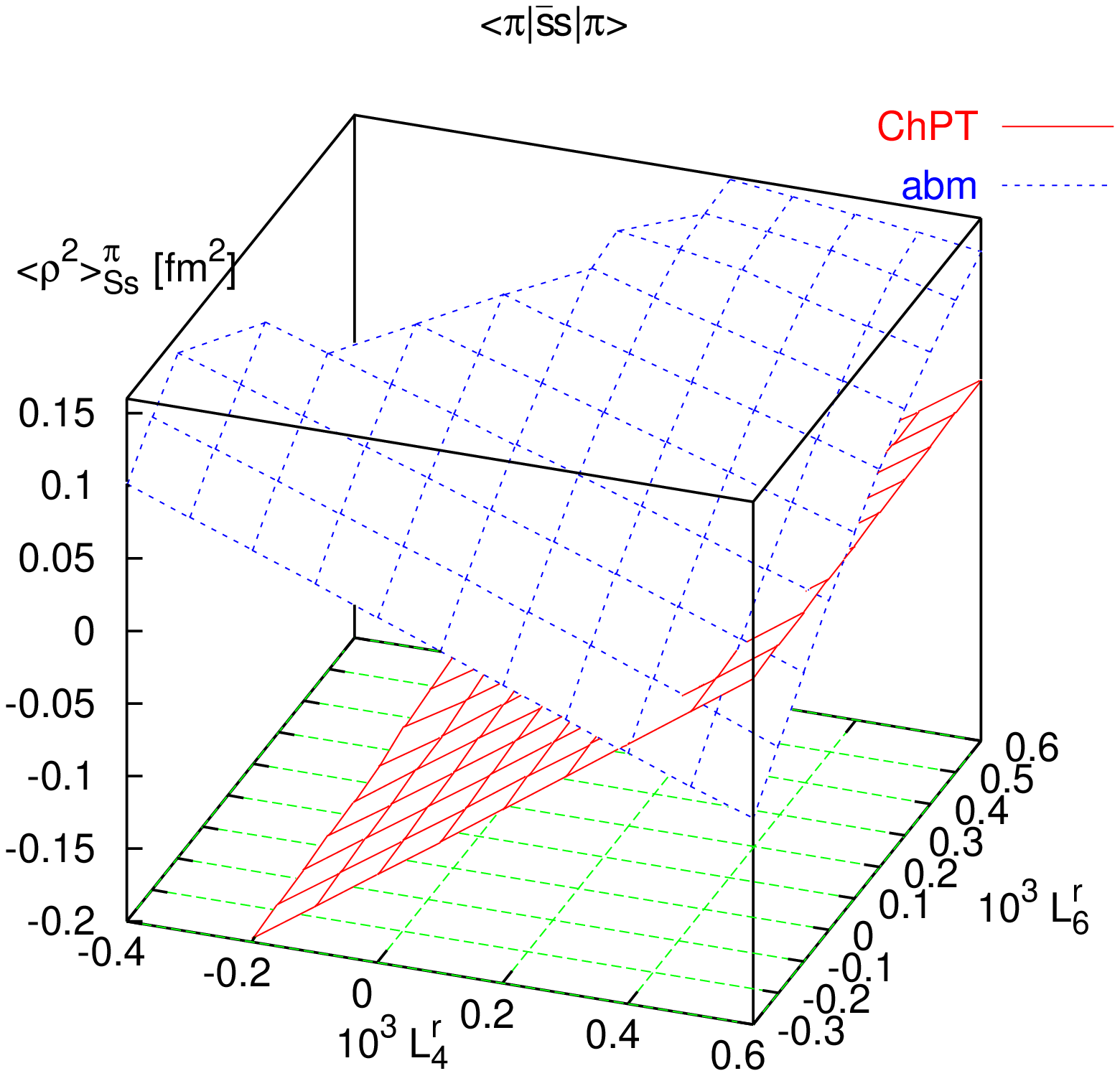}
\\(a)
\end{center}
\end{minipage}
\begin{minipage}{0.475\textwidth}
\begin{center}
\includegraphics[height=0.99\textwidth,angle=-90]{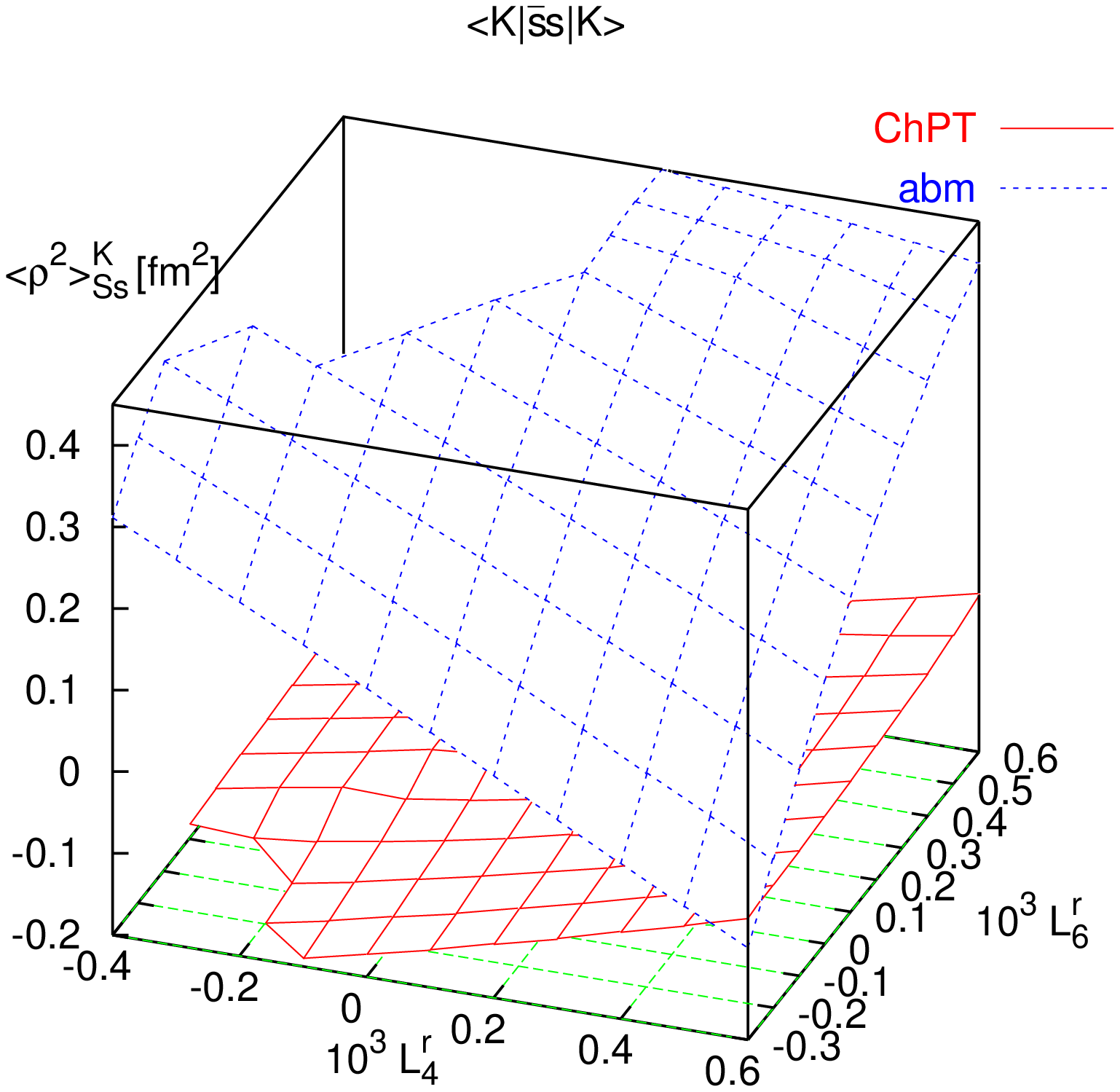}
\\(b)
\end{center}
\end{minipage}
\caption{ (a) The result for the pion strange quark scalar radius 
$<\rho^2>_{\pi s}^S$
as a function of  $L_4^r$ and $L_6^r$, with the dispersive estimate (abm)
and the ChPT calculation. (b) The result for the kaon strange quark scalar 
radius 
$<\rho^2>_{K s}^S$
as a function of  $L_4^r$ and $L_6^r$, with the dispersive estimate (abm)
and the ChPT calculation.}
\label{figrpis2}
\end{figure}

The strange quark scalar current can be studied in exactly the same way.
The pion strange scalar form factor
has a very small value at zero, we normalize therefore to
the value $B_0$ and plot
\be
<\rho^2>_{S s}^\pi  =  \frac{6}{B_0}\d{t} F^\pi_{Ss}(t)|_{t=0}\,.
\ee
The results are shown in Fig.~\ref{figrpis2}a both for the
dispersive and the ChPT result. Again the dispersive result shows a significant
variation with $L_4^r$ and $L_6^r$ since both solutions of the MO equations
are important and their contributions vary significantly depending on the
values of  $L_4^r$ and $L_6^r$.

The prediction for the kaon strange quark radius
\be
<r^2>_{Ss}^K  =  \frac{6}{ F^K_{Ss}(0)}\d{t} F^K_{Ss}(t)|_{t=0}\,,
\ee
is stable for most values of the $L_4^r$, $L_6^r$ except for a region
near the top of the $L_4^r$ and the bottom of the $L_6^r$ range
where $F^K_{Ss}(0)$ is near zero. The reason is that for most input values
the second MO solution dominates.
The two predictions never agree, the dispersive estimate is positive
and the ChPT one is negative.
We plot instead
\be
<\rho^2>_{Ss}^K  =  \frac{6}{B_0}\d{t} F^K_{Ss}(t)|_{t=0}\,,
\ee
in Fig.~\ref{figrpis2}b. For this quantity there
is a small region of overlap between the dispersive and the ChPT
estimates but it is where $F^K_{Ss}(0)$ is near zero and we can have
a large sensitivity to effects from the $C_i^r$.

It is somewhat difficult to get a final conclusion about $L_4^r$ and $L_6^r$
since the effect of the $p^6$ constants has to be evaluated. This requires
a general study of effects in the scalar sector which goes beyond the scope
of this paper.\footnote{The analysis might be doable by extending
some of the lines of work of \cite{EGPR}.} 
The main constraint from the pion scalar radius is
\be
L_6^r\approx L_4^r -0.00035\,.
\ee
To be precise, this is the curve where the two surfaces shown in
Fig.~\ref{figrpi2}a intersect.
If in addition we require that the values of the scalar form factors at zero
do not deviate too much from their lowest order values
we obtain that
$L_4^r$ should be in the range from $0.0003$ to $0.0006$. The latter
requirement means staying close to the intersection of the two surfaces
in Fig.~\ref{figF10ud}a.

We have shown
several numerical results in Table~\ref{tab:results} for grid points in this
region.

\subsection{Curvatures and the value of $C_{12}^r$ and $C_{13}^r$}
\label{c12c13}

Since the values at zero determine the form factors away from zero, we
can also compare higher order terms in the expansion in $t$.
Because of the large corrections to the values at zero coming from order
$p^6$, there is a rather large uncertainty in the resulting
values. We have therefore not done a full error analysis.

\begin{figure}
\begin{minipage}{0.475\textwidth}
\begin{center}
\includegraphics[height=0.99\textwidth,angle=-90]{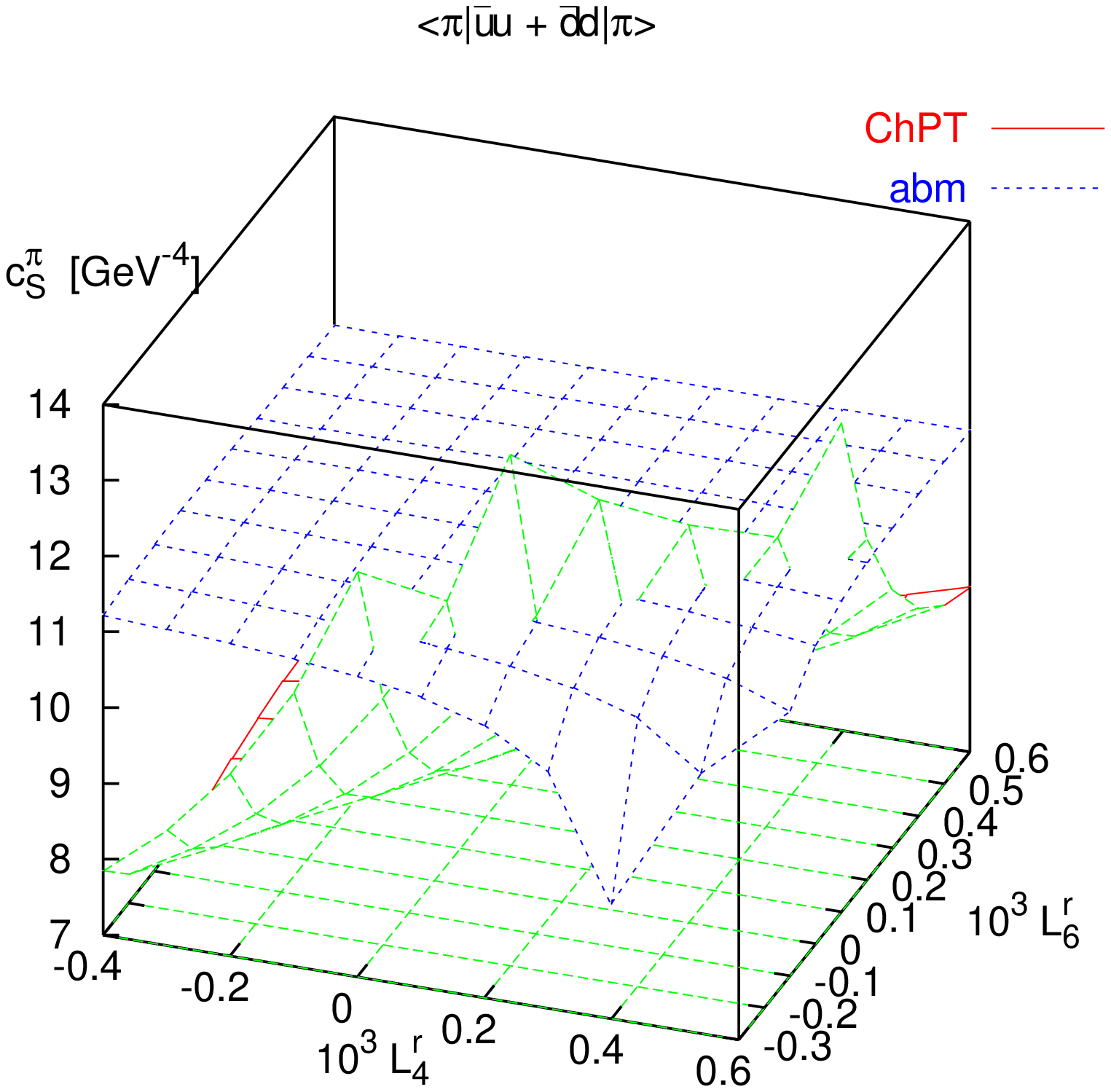}
\\(a)
\end{center}
\end{minipage}
\begin{minipage}{0.475\textwidth}
\begin{center}
\includegraphics[height=0.99\textwidth,angle=-90]{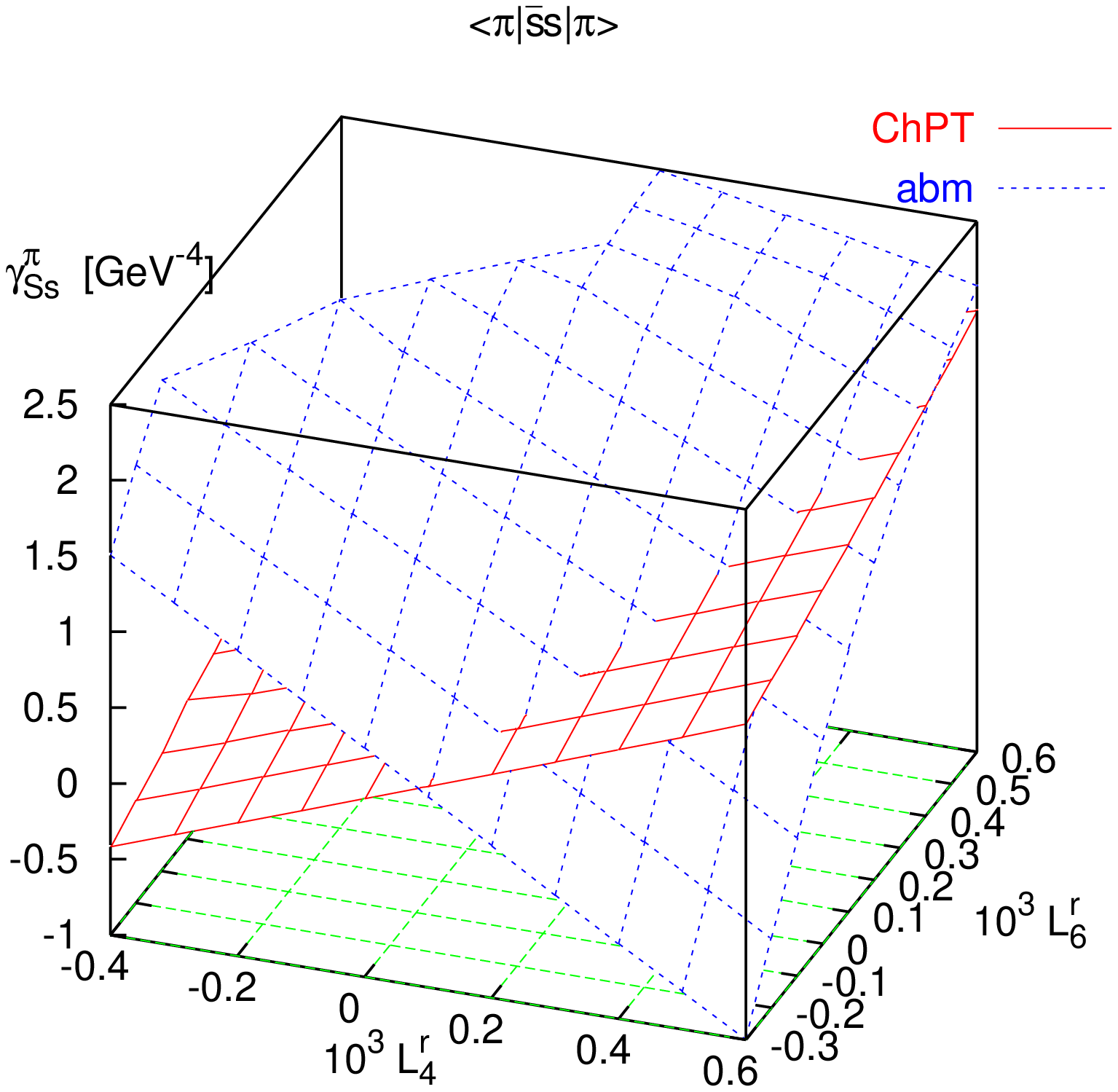}
\\(b)
\end{center}
\end{minipage}
\caption{(a) The result for the pion scalar curvature
$c^\pi_S$
as a function of  $L_4^r$ and $L_6^r$, with the dispersive estimate (abm)
and the ChPT calculation.
(b) The result for the pion strange scalar curvature
$\gamma^\pi_{Ss}$
as a function of  $L_4^r$ and $L_6^r$, with the dispersive estimate (abm)
and the ChPT calculation.}
\label{figcpi}
\end{figure}

We define the pion scalar curvature via
\be
c^\pi_S = \frac{1}{F_S^\pi(0)}\frac{1}{2}\dd{t} F_{S}^\pi(t)|_{t=0}\,.
\ee
The results both from the dispersive and the ChPT analysis are shown in
Fig.~\ref{figcpi}a.
It should be remarked that, precisely as in the two flavour case analyzed in
\cite{BCT} a large part of the curvature is due to the loop effects
and the contribution from the $p^6$ constants is not the dominant one.
This can be seen by looking at the scale in Fig.~\ref{figcpi}a. The difference
between the ChPT prediction with the $C_i^r=0$
and the dispersive result is small.

The same analysis can be performed for the other form factors. We again run
into the problem of normalizing the curvature, just as we had for the slopes.
For the rest we use thus quantities normalized to $B_0$.
We show the pion strange scalar curvature
\be
\label{gammapis}
\gamma^\pi_{Ss} = 
 \frac{1}{B_0}\frac{1}{2}\dd{t} F_{Ss}^\pi(t)|_{t=0}\,.
\ee
in Fig.~\ref{figcpi}b. The variation of the dispersive result is large here
again due to the fact that both MO solutions contribute with a strong variation
in relative strength. 

Analogously to (\ref{gammapis}) we define
\ba
\gamma^\pi_{S} &=& 
 \frac{1}{2B_0}\frac{1}{2}\dd{t} F_{S}^\pi(t)|_{t=0}\,,
\nonumber\\
\gamma^K_{Sq} &=& 
 \frac{1}{B_0}\frac{1}{2}\dd{t} F_{Sq}^K(t)|_{t=0}\,,
\nonumber\\
\gamma^K_{S} &=& 
 \frac{1}{B_0}\frac{1}{2}\dd{t} F_{Ss}^K(t)|_{t=0}\,.
\ea
but we have not included plots of these quantities.

The curvatures can now be used to estimate combinations of the $C_i^r$.
We get 
\ba
C_{12}^r+2C_{13}^r &=& -\frac{F_\pi^4}{8}\left( c_S^{\pi\, disp} 
- c_S^{\pi\, ChPT, C_i^r=0}\right)
 = -\frac{F_\pi^4}{8}\left( \gamma_{Ss}^{K\, disp} 
- \gamma_{Ss}^{K\, ChPT, C_i^r=0}\right)\,,
\nonumber\\
C_{13}^r &=& -\frac{F_\pi^4}{16}\left( \gamma_{Ss}^{\pi disp}-
\gamma_{Ss}^{\pi ChPT, C_i^r=0} \right)\,,
\nonumber\\
C_{12}^r+4 C_{13}^r &=& -\frac{F_\pi^4}{8}\left( \gamma_{Sq}^{K\, disp} 
- \gamma_{Sq}^{K\, ChPT, C_i^r=0}\right)
\ea
From the figures \ref{figcpi}a and  \ref{figcpi}b
it is obvious that the extracted values will
depend strongly on the input values used for $L_4^r$ and $L_6^r$, since the
difference between the surfaces in those figures gives the values of the
$C_i^r$.
On the other hand, we see that there are several ways to extract
the same quantities. It turns out that these tend to give similar values
in the region of $L_i^r$ preferred in the previous subsections.
Away from these regions, the obtained values for the $C_i^r$ are
much larger than naive order of magnitude estimates.

\begin{table}
\begin{center}
\begin{tabular}{ccccc}
\hline
\rule{0cm}{2.6ex}Set            & Fit 10 & A & B & C \\[0.25ex]
\hline
\rule{0cm}{2.6ex}$10^3\cdot L_4^r$                   & 0.0 & 0.4 & 0.5 & 0.5 \\
$10^3\cdot L_6^r$                   & 0.0 & 0.1 & 0.1 & 0.2 \\[0.25ex]
\hline
\rule{0cm}{2.6ex}$F_S^\pi(0)/B_0$ (ChPT, $C_i^r=0$)& 2.54 & 1.99 & 1.75 &2.12\\
$F_{Ss}^\pi(0)/B_0$    (ChPT, $C_i^r=0$)& 0.020 & 0.004 & $-$0.002 & 0.008\\
$F_{Sq}^K(0)/B_0$    (ChPT, $C_i^r=0$)  & 1.94 & 1.36 & 1.12 & 1.51 \\
$F_{Ss}^K(0)/B_0$   (ChPT, $C_i^r=0$)   & 1.77 & 1.35 & 1.17 & 1.45 \\[0.25ex]
\hline
\rule{0cm}{2.6ex}$<r^2>^{\pi, disp}_S$ (fm$^2$)     
     & 0.617 &  0.612 &  0.610 & 0.614\\
$<r^2>^{\pi, ChPT, C_i^r=0}_S$ (fm$^2$) & 0.384 &  0.547 &  0.625 & 0.563\\
\hline
\rule{0cm}{2.6ex}$10^5 \left(C_{12}^r+2C_{13}^r\right)\, [c_S^\pi]$
 & $-$2.6 &  $-$0.56 &   0.55 & $-$0.71 \\
$10^5 \left(C_{12}^r+2C_{13}^r\right) [\gamma_S^\pi]$
 & $-$3.3 &  $-$0.55 &   0.48 & $-$0.75 \\
$10^5 \left(C_{12}^r+2C_{13}^r\right) [\gamma_{Ss}^K]$
 & $-$0.55 &    0.11 &   0.33 &    0.15 \\
$10^5\,C_{13}^r [\gamma_{Ss}^\pi]$
 & $-$0.56&  $-$0.02 &   0.15 &    0.03 \\
$10^5\left(C_{12}^r+4C_{13}^r\right) [\gamma_{Sq}^K]$
 & $-$4.1 &  $-$0.27 &   0.99  &$-$0.08 \\
$10^5\,C_{12}^r [c_S^\pi ,\,\gamma_{Ss}^\pi]$ &
 $ -$1.5&    $-$0.52 &   0.26 & $-$0.78    \\
$10^5\,C_{12}^r [c_S^\pi ,\,\gamma_{Sq}^K]$ &
 $-$1.1 &    $-$0.84 &   0.12 & $-$1.3   \\[0.25ex]
\hline
\rule{0cm}{2.6ex}
$F_0$ (MeV)        & 87.7      & 63.5  & 70.4  & 71.0 \\
$ F_{\pi^+}^{(4)}$ & 0.136     &0.230  &0.253  &0.254 \\
 $ F_{\pi^+}^{(6)}$& $-$0.083  &0.226  &0.059  &0.048 \\
$ F_{K\pi}^{(4)} $ & 0.169     &0.157  &0.153  &0.159 \\
$  F_{K\pi}^{(6)}$ & 0.051     &0.063  &0.067  &0.061 \\[0.25ex]
\hline
$ \mu_\pi^{(2)}  $ & 0.736     &1.005   &1.129   &0.936 \\
$\mu_\pi^{(4)}   $ & 0.006     &$-$0.090&$-$0.138&$-$0.043\\
$ \mu_\pi^{(6)}  $ & 0.258     &0.085   &0.009   &0.107 \\
$\mu_K^{(2)}     $ & 0.687     &0.938   &1.055   &0.874 \\
$\mu_K^{(4)}     $ & 0.007     &$-$0.100&$-$0.149&$-$0.057\\
$\mu_K^{(6)}     $ & 0.306     &0.162   &0.094   &0.183 \\
$\mu_\eta^{(2)}  $ & 0.734     &1.001   &1.124   &0.933 \\
$\mu_\eta^{(4)}  $ & $-$0.052  &$-$0.151&$-$0.197&$-$0.104 \\
$\mu_\eta^{(6)}  $ & 0.318     &0.150   &0.073   &0.171 \\[0.25ex]
\hline
\end{tabular}
\end{center}
\caption{The results for various quantities for the $L_i^r$ of fit 10
and sets A, B and C. These are described in the text.}
\label{tab:results}
\end{table}

We have collected in Table~\ref{tab:results} various
results for four sets of $L_i^r$. The first one is exactly the same
as fit 10 of Ref.~\cite{ABT4} and all quantities coincide with the
one reported there.
The notation $C_{12}^r+2C_{13}^r [c_S^\pi]$ used in Table~\ref{tab:results}
means that the
quantity  $C_{12}^r+2C_{13}^r$ is determined from $c^\pi_S$ using the
$L_i^r$ as determined from a fit to the same experimental
input as used in fit 10 in Ref.~\cite{ABT4} but with the values of $L_4^r$
and $L_6^r$ as indicated. The difference between the first two determinations
of $C_{12}^r+2C_{13}^r$ is due to the difference from $F^\pi_S(0)$ from
$2B_0$.

The three other sets - A,B, and C - of $L_i^r$ were
generated by different sets of $L_4^r$ and $L_6^r$ in the agreement 
region of the two previous sections, $L_6^r\approx L_4^r -0.00035$ 
and \mbox{$0.0003 \lesssim L_4^r \lesssim 0.0006$.} The extracted
constants are generally similar over the three sets but still show 
a dependence on $L_4^r$ and $L_6^r$. There are particular disagreements
on values obtained via $\gamma_{Sq}^K$ and $\gamma_{Ss}^{\pi}$. For the 
former, we can, as we did for the light quark scalar radius, attribute 
the variations over the extracted values to the construction of 
$\gamma_{Sq}^K$ on the canonical MO solutions. For $\gamma_{Ss}^{\pi}$,
the argument is similar, as now the relative strength of the MO solutions
is compensated by the values of $F_{Ss}^\pi(0)/B_0$ and $F_{Ss}^K(0)/B_0~$.

A final conclusion on the value of $C_{12}^r$ which is needed
for the measurement of $V_{us}$ \cite{BT2} is not very easy since
it depends so strongly on the values of $L_4^r$ and $L_6^r$ as well as 
on which dispersive quantities are used to evaluate it.

\subsection{Masses and Decay Constants}
\label{Masses}

For completeness we present here results for the masses and decay
constants for the sets of input parameters we use.
These can also be found in Table~\ref{tab:results}. We present results
for
\ba
F_{\pi^+}/F_0  &=& 1 +  F_{\pi^+}^{(4)} +  F_{\pi^+}^{(6)}\,,
\nonumber\\
F_{K^+}/F_{\pi^+}  &=& 1 +  F_{K\pi}^{(4)} +  F_{K\pi}^{(6)}\,,
\nonumber\\
m_{\pi^\pm}^2/(m_{\pi^\pm}^2)_{\mbox{\tiny QCD}}
& = & \mu_\pi^{(2)}+\mu_\pi^{(4)}+ \mu_\pi^{(6)}
\nonumber\\
m_{K^\pm}^2/(m_{K^\pm}^2)_{\mbox{\tiny QCD}}
& = &\mu_K^{(2)}+\mu_K^{(4)}+ \mu_K^{(6)} \,,
\nonumber\\
m_{\eta}^2/(m_{\eta}^2)_{\mbox{\tiny phys}}
& = & \mu_\eta^{(2)}+\mu_\eta^{(4)}+ \mu_\eta^{(6)}\,.
\ea
Notice that in \cite{ABT4} the corresponding numbers were quoted for the
``Main Fit'' which used the old $K_{e4}$ data.

A glance at Table~\ref{tab:results} shows that the pion decay constant
in the chiral limit $F_0$ can be substantially different from the value
of about $87~$MeV for the case of fit 10.

We would like to remind the reader that from within $SU(3)$ ChPT
all the solutions shown in the table are good fits to all fitted
experiments. The new sets are preferred under the assumptions that
the $C_i^r$ contributing to the scalar form factors and masses vanish
as well as the constraints from assuming the corrections to the
scalar form factors at zero to be small.

\section{Conclusions}

In this paper we have computed the full order $p^6$ expressions in Chiral 
Perturbation Theory in the isospin limit for all pion and kaon scalar 
form factors.  This is described in the first major part of this paper.
In addition we have shown various numerical results that allow the reader
to get a feeling for the size of the various contributions.

We also found a new relation between the scalar form 
factors, analogous to the Sirlin relation for vector form factors.

This calculation is important for future determination of
the $p^6$ LEC's and in the absence of a better knowledge of these constants
the precision on numerical results is limited.

In the second major part of this paper we have used the experimental input
used in the previous fits of the order $p^4$ LECs and refitted the $L_i^r$
as a function of the input values of $L_4^r$ and $L_6^r$ showing that
a rather wide range of these leads to good fits to the included data.
We also used the solutions of the 
Muskhelishvili-Omn\`es problem for the scalar form factors
with as input the known $\pi\pi$ and $KK$ scattering
phases. This allowed to study the scalar radii and curvatures as a function
of $L_4^r$ and $L_6^r$ and obtain consistency relations between them.
Under the assumptions of vanishing $C_i^r$ contributions to the
scalar form factors at $t=0$ and the scalar radii these consistency
requirements together with the assumption that the corrections to the
scalar form factors at zero should be small, there is a preferred region
of $L_4^r$ and $L_6^r$ around $L_4^r\approx 0.00045$ and
$L_6^r\approx 0.00015$. In Table \ref{tab:results} we have shown the
values of several quantities in this region.

We have used our results for the curvature to obtain constraints on
two order $p^6$ constants, $C_{12}^r$ and $C_{13}^r$. These are
of the expected naive order of magnitude but the precision obtained
is rather low.

\section*{Acknowledgements}
This work has been funded in part by
the Swedish Research Council and the European Union RTN
network, Contract No. HPRN-CT-2002-00311  (EURIDICE).
FORM 3.0 has been used extensively in these calculations \cite{FORM3}.
We thank B. Moussallam for providing us with the programs
of Refs.~\cite{Moussallam1,ABM}.

\appendix
\renewcommand{\theequation}{\Alph{section}.\arabic{equation}}
\setcounter{equation}{0}

\section{The proof of the symmetry relation}
\label{app:relation}

The relation for the scalar form factors, Eq.~(\ref{eq:relation}) can be
proven using the same method as used in the second paper of Ref.~\cite{Sirlin}
when the different behaviour of scalar and vector currents under charge
conjugation is taken in to account.

We write the Hamiltonian in the form
\be
H = H_0+\lambda_3 v_3\,.
\ee
$H_0$ is the Hamiltonian in the limit $m_s = m_u$. We now do oldfashioned
perturbation theory in $\lambda_3$, and look at the variation of
\be
\label{Kpi1}
F_{su}^{K^0\pi^-}(t)
\ee
w.r.t. $\lambda_3$ at $\lambda_3=0$. This can be rewritten as
an expectation value involving the $K^0$ and $\pi^-$ field, $v_3$
and $\overline{s}u$. Applying the operator $G_V = C exp(i\pi V_2)$ with $V_2$
the second generator of $V$-spin and $C$ charge conjugation we obtain that
\be
\label{Kpirelation}
\left.\frac{\partial}{\partial\lambda_3}
F_{su}^{K^0\pi^-}(t)\right|_{\lambda_3=0}
= 
-\left.\frac{\partial}{\partial\lambda_3}
F_{su}^{\overline{K^0}\pi^+}(t)\right|_{\lambda_3=0}
\ee
while charge conjugation and the fact that it is a scalar current
require the full form factors in (\ref{Kpirelation}) to be equal.
The first variation w.r.t. $\lambda_3$ of $F_{su}^{K^0\pi^-}(t)$
thus vanishes.

The same argument can be applied to the combination
\be
\label{Kpi2}
F_{uu}^{\pi^+\pi^+}(t)-F_{ss}^{\pi^+\pi^+}(t)
-F_{uu}^{\overline{K^0}\,\overline{K^0}}(t)
+F_{ss}^{\overline{K^0}\,\overline{K^0}}(t)\,.
\ee
and in the $V$-spin limit both combinations are related. This relation can be
brought into the form of Eq.~(\ref{eq:relation}) using isospin.

\section{Analytical expressions for $F_S^\pi(t)$}
\label{Appresults}

For brevity we only write the pion scalar form factor with the
light quark densities. The expressions for the others can be obtained
from the authors. The integrals can be found in several places.
The functions $\overline{A}$, $\overline{B}$,\ldots can be found
in \cite{ABT1} and \cite{ABT3}. The $H^F$,\ldots  are defined in
\cite{ABT1} and the $V_i$ can be found in \cite{BT1}. The method to evaluate
the $H^F_i$ was developed in \cite{ABT1} and the $V_i$ are evaluated
as described in \cite{BT1} using the methods of \cite{Ghinculov}.

The result given is not the shortest possible analytical one. There
are various relations between the integrals that we have not implemented.
The reason is that these relations involve inverse powers of $t$ and we
have used simplifying relations to rewrite all the masses in the numerators
in terms of $m_\pi^2$ and $m_K^2$. This has as a consequence that, if we
had used those relations, the numerators of the $1/t$ poles would not cancel
exactly numerically and produce possible instabilities for small $t$.

We write the pion scalar form factor as
\be
F_S^{\pi}(t) = 2 B_0 \left(F_S^{\pi(2)}+\frac{1}{F_\pi^2} F_S^{\pi(4)}(t)
+\frac{1}{F_\pi^4} F_S^{\pi(6)}(t)\right)\,.
\ee
The superscript indicates the chiral order. The lowest order is simply
\be
F_S^{\pi(2)} = 1\,.
\ee

The next order has been calculated in \cite{GL3} and we agree with their
result, it is
\ba
F_S^{\pi(4)}(t) &=&
        - 16 L_4^r m_K^2 - 24 L_4^r m_\pi^2 + 8 L_4^r t - 16 L_5^r m_\pi^2 + 
         4 L_5^r t + 32 L_6^r m_K^2 
\nonumber\\&&
+ 48 L_6^r m_\pi^2 + 32 L_8^r m_\pi^2 
         + 1/6 \overline{A}(m_\eta^2) - 1/2 \overline{A}(m_\pi^2) 
       + \overline{B}(m_\eta^2,m_\eta^2,t)  ( 1/18 m_\pi^2 )
\nonumber\\&&
       + \overline{B}(m_K^2,m_K^2,t)  ( 1/4 t )
       + \overline{B}(m_\pi^2,m_\pi^2,t)  (  - 1/2 m_\pi^2 + t )\,.
\ea

The next order we split in several different parts
\be
F_S^{\pi(6)}(t) = F_{S\mathbf{B}}^{\pi}(t)+F_{S\mathbf{H}}^{\pi}(t)
 +F_{S\mathbf{V}}^{\pi}(t)+F_{S\mathbf{L}}^{\pi}(t)+F_{S\mathbf{C}}^{\pi}(t)\,.
\ee
The result for $F_{S\mathbf{C}}^{\pi}(t)$ can also be found in the main text
we simply repeat it here for completeness.
\ba
F_{S\mathbf{C}}^{\pi}(t) &=&
        - 128 m_\pi^2 m_K^2 C_{13}^r - 64 m_\pi^2 m_K^2 C_{15}^r
 + 64 m_\pi^2 m_K^2 C_{16}^r + 64
          m_\pi^2 m_K^2 C_{20}^r
\nonumber\\&&
 + 576 m_\pi^2 m_K^2 C_{21}^r + 128 m_\pi^2 m_K^2 C_{32}^r 
+ 32 m_\pi^2 C_{12}^r t
          + 64 m_\pi^2 C_{13}^r t + 16 m_\pi^2 C_{14}^r t
\nonumber\\&&
 + 24 m_\pi^2 C_{15}^r t 
+ 32 m_\pi^2 C_{16}^r 
         t + 16 m_\pi^2 C_{17}^r t + 16 m_\pi^2 C_{34}^r t
 + 16 m_\pi^2 C_{36}^r t 
- 96 m_\pi^4 
         C_{12}^r
\nonumber\\&&
 - 128 m_\pi^4 C_{13}^r - 48 m_\pi^4 C_{14}^r 
- 64 m_\pi^4 C_{15}^r - 112 m_\pi^4 
         C_{16}^r - 48 m_\pi^4 C_{17}^r + 144 m_\pi^4 C_{19}^r 
\nonumber\\&&
+ 240 m_\pi^4 C_{20}^r + 240 m_\pi^4 
         C_{21}^r + 96 m_\pi^4 C_{31}^r + 128 m_\pi^4 C_{32}^r 
+ 16 m_K^2 C_{15}^r t - 64 m_K^4 
         C_{16}^r
\nonumber\\&&
 + 64 m_K^4 C_{20}^r + 192 m_K^4 C_{21}^r 
- 8 C_{12}^r t^2 - 16 C_{13}^r t^2
        \,.
\ea

\ba
   F_{S\mathbf{H}}^\pi(t) &=&
       H^{F\prime}(m_\pi^2,m_\pi^2,m_\pi^2,m_\pi^2)   ( 5/6 m_\pi^4 )
       + H^{F\prime}(m_\pi^2,m_K^2,m_K^2,m_\pi^2)   (  - 5/8 m_\pi^4 )
\nonumber\\&&
       + H^{F\prime}(m_\pi^2,m_\eta^2,m_\eta^2,m_\pi^2)   ( 1/18 m_\pi^4 )
       + H^{F\prime}(m_K^2,m_\pi^2,m_K^2,m_\pi^2)   ( m_\pi^2 m_K^2 )
\nonumber\\&&
       + H^{F\prime}(m_K^2,m_K^2,m_\eta^2,m_\pi^2)   (  - 5/6 m_\pi^4 )
\nonumber\\&&
       + H^{F\prime}(m_\eta^2,m_K^2,m_K^2,m_\pi^2)  
 ( 1/2 m_\pi^2 m_K^2 + 7/8 m_\pi^4 )
\nonumber\\&&
       + H_{1}^{F\prime}(m_\pi^2,m_K^2,m_K^2,m_\pi^2)   ( m_\pi^4 )
       + H_{1}^{F\prime}(m_\eta^2,m_K^2,m_K^2,m_\pi^2)   (  - m_\pi^4 )
\nonumber\\&&
       + H_{21}^{F\prime}(m_\pi^2,m_\pi^2,m_\pi^2,m_\pi^2)   ( 3 m_\pi^4 )
       + H_{21}^{F\prime}(m_\pi^2,m_K^2,m_K^2,m_\pi^2)   (  - 3/8 m_\pi^4 )
\nonumber\\&&
       + H_{21}^{F\prime}(m_K^2,m_\pi^2,m_K^2,m_\pi^2)   ( 3 m_\pi^4 )
       + H_{21}^{F\prime}(m_\eta^2,m_K^2,m_K^2,m_\pi^2)   ( 9/8 m_\pi^4 )
\nonumber\\&&
       + H^F(m_\pi^2,m_\pi^2,m_\pi^2,m_\pi^2)   ( 2/3 m_\pi^2 - 2/3 t )
\nonumber\\&&
       + H^F(m_\pi^2,m_K^2,m_K^2,m_\pi^2)   (  - 47/24 m_\pi^2 + 1/6 t )
       + H^F(m_\pi^2,m_\eta^2,m_\eta^2,m_\pi^2)   ( 1/9 m_\pi^2 )
\nonumber\\&&
       + H^F(m_K^2,m_\pi^2,m_K^2,m_\pi^2)   ( 5/6 m_\pi^2 + m_K^2 - 1/2 t )
\nonumber\\&&
       + H^F(m_K^2,m_K^2,m_\eta^2,m_\pi^2)   (  - 13/9 m_\pi^2 - 1/8 t )
\nonumber\\&&
       + H^F(m_\eta^2,m_K^2,m_K^2,m_\pi^2)   ( 119/72 m_\pi^2 + 1/2 m_K^2 
             + 1/24 t )
\nonumber\\&&
       + H_{1}^F(m_\pi^2,m_K^2,m_K^2,m_\pi^2)   ( 2 m_\pi^2 + 1/24 t )
\nonumber\\&&
       + H_{1}^F(m_\eta^2,m_K^2,m_K^2,m_\pi^2)   (  - 2 m_\pi^2 - 1/24 t )
\nonumber\\&&
       + H_{21}^F(m_\pi^2,m_\pi^2,m_\pi^2,m_\pi^2)   ( 6 m_\pi^2 )
       + H_{21}^F(m_\pi^2,m_K^2,m_K^2,m_\pi^2)   (  - 3/4 m_\pi^2 )
\nonumber\\&&
       + H_{21}^F(m_K^2,m_\pi^2,m_K^2,m_\pi^2)   ( 6 m_\pi^2 )
       + H_{21}^F(m_\eta^2,m_K^2,m_K^2,m_\pi^2)   ( 9/4 m_\pi^2 )\,.
\ea

\ba
   F_{S\mathbf{V}}^\pi(t) &=&
       + V_{0}(m_\pi^2,m_\pi^2,m_\pi^2,m_\pi^2,m_\pi^2,t,m_\pi^2) 
             (  - m_\pi^2 t + 7/2 m_\pi^4 + 2/
         3 t^2 )
\nonumber\\&&
       + V_{0}(m_\pi^2,m_\pi^2,m_K^2,m_K^2,m_\pi^2,t,m_\pi^2)  
         (  - 11/12 m_\pi^2 t + 3/2 
         m_\pi^4 + 1/12 t^2 )
\nonumber\\&&
       + V_{0}(m_\pi^2,m_\pi^2,m_\eta^2,m_\eta^2,m_\pi^2,t,m_\pi^2)  
           ( 1/18 m_\pi^4 )
\nonumber\\&&
       + V_{0}(m_\pi^2,m_\eta^2,m_K^2,m_K^2,m_\pi^2,t,m_\pi^2) 
          ( 1/4 m_\pi^2 t - 1/6 m_\pi^4 )
\nonumber\\&&
       + V_{0}(m_K^2,m_K^2,m_\pi^2,m_K^2,m_\pi^2,t,m_\pi^2) 
            (  - 17/12 m_\pi^2 t + 3/2 
         m_\pi^4 + 1/2 t^2 )
   \nonumber\\&& + V_{0}(m_K^2,m_K^2,m_K^2,m_\eta^2,m_\pi^2,t,m_\pi^2)  
           (  - 2/3 m_\pi^2 t + 2/3 m_\pi^4
          + 1/8 t^2 )
   \nonumber\\&& + V_{0}(m_\eta^2,m_\pi^2,m_K^2,m_K^2,m_\pi^2,t,m_\pi^2) 
          ( 2/9 m_\pi^2 m_K^2 + 13/36 m_\pi^2  t
 \nonumber\\&& ~~~~~~ 
        - 7/18 m_\pi^4 - 1/9 m_K^2 t - 1/12 t^2 )
   \nonumber\\&& + V_{0}(m_\eta^2,m_\eta^2,m_\pi^2,m_\eta^2,m_\pi^2,t,m_\pi^2) 
          ( 1/27 m_\pi^4 )
   \nonumber\\&& + V_{0}(m_\eta^2,m_\eta^2,m_K^2,m_K^2,m_\pi^2,t,m_\pi^2)
             (  - 5/36 m_\pi^2 t + 1/18 
         m_\pi^4 + 1/12 t^2 )
   \nonumber\\&& + V_{11}(m_\pi^2,m_\pi^2,m_\pi^2,m_\pi^2,m_\pi^2,t,m_\pi^2) 
             ( 13/3 m_\pi^2 t - 4 m_\pi^4 - 
         4/3 t^2 )
   \nonumber\\&& + V_{11}(m_\pi^2,m_\pi^2,m_K^2,m_K^2,m_\pi^2,t,m_\pi^2)  
            ( 11/6 m_\pi^2 t - 3 m_\pi^4 - 
         1/6 t^2 )
   \nonumber\\&& + V_{11}(m_\pi^2,m_\eta^2,m_K^2,m_K^2,m_\pi^2,t,m_\pi^2) 
             (  - 1/2 m_\pi^2 t + 2/3 
         m_\pi^4 )
   \nonumber\\&& + V_{11}(m_K^2,m_K^2,m_\pi^2,m_K^2,m_\pi^2,t,m_\pi^2) 
             ( 53/12 m_\pi^2 t - 4 m_\pi^4
          - t^2 )
   \nonumber\\&& + V_{11}(m_K^2,m_K^2,m_K^2,m_\eta^2,m_\pi^2,t,m_\pi^2)  
          ( 5/3 m_\pi^2 t - 2 m_\pi^4 - 1/
         4 t^2 )
   \nonumber\\&& + V_{11}(m_\eta^2,m_\pi^2,m_K^2,m_K^2,m_\pi^2,t,m_\pi^2)  
            (  - 2/9 m_\pi^2 m_K^2 - 7/9 m_\pi^2 t
\nonumber\\&& ~~~~~~     
     + 8/9 m_\pi^4 + 1/9 m_K^2 t + 1/6 t^2 )
   \nonumber\\&&
  + V_{11}(m_\eta^2,m_\eta^2,m_K^2,m_K^2,m_\pi^2,t,m_\pi^2) 
               ( 5/9 m_\pi^2 t - 1/3 m_\pi^4
          - 1/6 t^2 )
   \nonumber\\&& + V_{12}(m_\pi^2,m_\pi^2,m_\pi^2,m_\pi^2,m_\pi^2,t,m_\pi^2)
              (  - t^2 )
   \nonumber\\&& + V_{12}(m_\pi^2,m_\pi^2,m_K^2,m_K^2,m_\pi^2,t,m_\pi^2) 
            ( 1/3 m_\pi^2 t + 1/8 t^2 )
   \nonumber\\&& + V_{12}(m_\pi^2,m_\eta^2,m_K^2,m_K^2,m_\pi^2,t,m_\pi^2) 
                (  - 1/6 m_\pi^2 t - 1/8 
         t^2 )
   \nonumber\\&& + V_{12}(m_K^2,m_K^2,m_\pi^2,m_K^2,m_\pi^2,t,m_\pi^2) 
                 ( 5/6 m_\pi^2 t - 1/2 t^2 )
   \nonumber\\&& + V_{12}(m_K^2,m_K^2,m_K^2,m_\eta^2,m_\pi^2,t,m_\pi^2)  
              ( 1/3 m_\pi^2 t )
   \nonumber\\&& + V_{12}(m_\eta^2,m_\pi^2,m_K^2,m_K^2,m_\pi^2,t,m_\pi^2)  
             (  - 5/18 m_\pi^2 t + 1/9 m_K^2
          t + 1/8 t^2 )
   \nonumber\\&& + V_{12}(m_\eta^2,m_\eta^2,m_K^2,m_K^2,m_\pi^2,t,m_\pi^2)  
           ( 1/9 m_\pi^2 t - 1/8 t^2 )
   \nonumber\\&& + V_{13}(m_\pi^2,m_\pi^2,m_\pi^2,m_\pi^2,m_\pi^2,t,m_\pi^2)  
            ( 2/3 m_\pi^2 t - 8 m_\pi^4 + 2/
         3 t^2 )
   \nonumber\\&& + V_{13}(m_\pi^2,m_\pi^2,m_K^2,m_K^2,m_\pi^2,t,m_\pi^2)  
            ( 2 m_\pi^2 t - 4 m_\pi^4 )
   \nonumber\\&& + V_{13}(m_\pi^2,m_\eta^2,m_K^2,m_K^2,m_\pi^2,t,m_\pi^2)  
               (  - 1/2 m_\pi^2 t + 1/3 
         m_\pi^4 )
   \nonumber\\&& + V_{13}(m_K^2,m_K^2,m_\pi^2,m_K^2,m_\pi^2,t,m_\pi^2) 
               ( 13/6 m_\pi^2 t - 3 m_\pi^4 - 
         5/24 t^2 )
   \nonumber\\&& + V_{13}(m_K^2,m_K^2,m_K^2,m_\eta^2,m_\pi^2,t,m_\pi^2)  
              ( 3/2 m_\pi^2 t - 2 m_\pi^4 - 1/
         8 t^2 )
   \nonumber\\&& + V_{13}(m_\eta^2,m_\pi^2,m_K^2,m_K^2,m_\pi^2,t,m_\pi^2)  
                (  - 4/9 m_\pi^2 m_K^2 - 13/18 
         m_\pi^2 t + 7/9 m_\pi^4
\nonumber\\&& ~~~~~~ + 2/9 m_K^2 t + 1/6 t^2 )
   \nonumber\\&& + V_{14}(m_\pi^2,m_\pi^2,m_\pi^2,m_\pi^2,m_\pi^2,t,m_\pi^2)  
                    ( 2/3 t^2 )
   \nonumber\\&& + V_{14}(m_\pi^2,m_\pi^2,m_K^2,m_K^2,m_\pi^2,t,m_\pi^2) 
              ( 1/2 t^2 )
   \nonumber\\&& + V_{14}(m_\pi^2,m_\eta^2,m_K^2,m_K^2,m_\pi^2,t,m_\pi^2) 
            ( 1/6 m_\pi^2 t - 1/4 t^2 )
   \nonumber\\&& + V_{14}(m_K^2,m_K^2,m_\pi^2,m_K^2,m_\pi^2,t,m_\pi^2) 
             ( 7/24 t^2 )
   \nonumber\\&& + V_{14}(m_K^2,m_K^2,m_K^2,m_\eta^2,m_\pi^2,t,m_\pi^2) 
               ( 1/4 t^2 )
   \nonumber\\&& + V_{14}(m_\eta^2,m_\pi^2,m_K^2,m_K^2,m_\pi^2,t,m_\pi^2) 
              (  - 7/18 m_\pi^2 t + 2/9 m_K^2
          t + 1/6 t^2 )
   \nonumber\\&& + V_{21}(m_\pi^2,m_\pi^2,m_\pi^2,m_\pi^2,m_\pi^2,t,m_\pi^2) 
            ( 4 m_\pi^2 - 2/3 t )
   \nonumber\\&& + V_{21}(m_\pi^2,m_\pi^2,m_K^2,m_K^2,m_\pi^2,t,m_\pi^2)  
              ( 3/2 m_\pi^2 - 7/12 t )
   \nonumber\\&& + V_{21}(m_\pi^2,m_\eta^2,m_K^2,m_K^2,m_\pi^2,t,m_\pi^2) 
           (  - 1/2 m_\pi^2 + 1/4 t )
   \nonumber\\&& + V_{21}(m_K^2,m_K^2,m_\pi^2,m_K^2,m_\pi^2,t,m_\pi^2)  
            ( 4 m_\pi^2 - t )
   \nonumber\\&& + V_{21}(m_K^2,m_K^2,m_K^2,m_\eta^2,m_\pi^2,t,m_\pi^2)  
              ( 3/2 m_\pi^2 - 1/2 t )
   \nonumber\\&& + V_{21}(m_\eta^2,m_\pi^2,m_K^2,m_K^2,m_\pi^2,t,m_\pi^2) 
             (  - 1/2 m_\pi^2 + 1/12 t )
   \nonumber\\&& + V_{21}(m_\eta^2,m_\eta^2,m_K^2,m_K^2,m_\pi^2,t,m_\pi^2)  
               ( 1/2 m_\pi^2 - 1/12 t )
   \nonumber\\&& + V_{22}(m_\pi^2,m_\pi^2,m_\pi^2,m_\pi^2,m_\pi^2,t,m_\pi^2) 
              (  - 10/3 m_\pi^2 t + 4 m_\pi^4
          + 2/3 t^2 )
   \nonumber\\&& + V_{22}(m_\pi^2,m_\pi^2,m_K^2,m_K^2,m_\pi^2,t,m_\pi^2) 
             (  - 11/12 m_\pi^2 t + 3/2 
         m_\pi^4 + 1/12 t^2 )
   \nonumber\\&& + V_{22}(m_\pi^2,m_\eta^2,m_K^2,m_K^2,m_\pi^2,t,m_\pi^2)
               ( 1/4 m_\pi^2 t - 1/2 m_\pi^4 )
   \nonumber\\&& + V_{22}(m_K^2,m_K^2,m_\pi^2,m_K^2,m_\pi^2,t,m_\pi^2)  
               (  - 3 m_\pi^2 t + 4 m_\pi^4 + 
         1/2 t^2 )
   \nonumber\\&& + V_{22}(m_K^2,m_K^2,m_K^2,m_\eta^2,m_\pi^2,t,m_\pi^2)  
                       (  - m_\pi^2 t + 3/2 m_\pi^4 + 
         1/8 t^2 )
   \nonumber\\&& + V_{22}(m_\eta^2,m_\pi^2,m_K^2,m_K^2,m_\pi^2,t,m_\pi^2)  
               ( 5/12 m_\pi^2 t - 1/2 m_\pi^4
          - 1/12 t^2 )
   \nonumber\\&& + V_{22}(m_\eta^2,m_\eta^2,m_K^2,m_K^2,m_\pi^2,t,m_\pi^2)  
                   (  - 5/12 m_\pi^2 t + 1/2 
         m_\pi^4 + 1/12 t^2 )
   \nonumber\\&& + V_{23}(m_\pi^2,m_\pi^2,m_\pi^2,m_\pi^2,m_\pi^2,t,m_\pi^2)  
                 ( 1/3 t^2 )
   \nonumber\\&& + V_{23}(m_\pi^2,m_\pi^2,m_K^2,m_K^2,m_\pi^2,t,m_\pi^2)  
               (  - 5/24 t^2 )
   \nonumber\\&& + V_{23}(m_\pi^2,m_\eta^2,m_K^2,m_K^2,m_\pi^2,t,m_\pi^2)  
                 ( 1/8 t^2 )
   \nonumber\\&& + V_{23}(m_K^2,m_K^2,m_K^2,m_\eta^2,m_\pi^2,t,m_\pi^2)  
                   (  - 1/8 t^2 )
   \nonumber\\&& + V_{23}(m_\eta^2,m_\pi^2,m_K^2,m_K^2,m_\pi^2,t,m_\pi^2) 
               (  - 1/24 t^2 )
   \nonumber\\&& + V_{23}(m_\eta^2,m_\eta^2,m_K^2,m_K^2,m_\pi^2,t,m_\pi^2) 
               ( 1/24 t^2 )
   \nonumber\\&& + V_{24}(m_\pi^2,m_\pi^2,m_\pi^2,m_\pi^2,m_\pi^2,t,m_\pi^2) 
             (  - 8/3 m_\pi^2 t + t^2 )
   \nonumber\\&& + V_{24}(m_\pi^2,m_\pi^2,m_K^2,m_K^2,m_\pi^2,t,m_\pi^2)  
              (  - 1/3 m_\pi^2 t - 1/8 
         t^2 )
   \nonumber\\&& + V_{24}(m_\pi^2,m_\eta^2,m_K^2,m_K^2,m_\pi^2,t,m_\pi^2) 
                ( 1/8 t^2 )
   \nonumber\\&& + V_{24}(m_K^2,m_K^2,m_\pi^2,m_K^2,m_\pi^2,t,m_\pi^2) 
              (  - 2 m_\pi^2 t + 1/2 t^2
          )
   \nonumber\\&& + V_{24}(m_K^2,m_K^2,m_K^2,m_\eta^2,m_\pi^2,t,m_\pi^2) 
             (  - 1/2 m_\pi^2 t )
   \nonumber\\&& + V_{24}(m_\eta^2,m_\pi^2,m_K^2,m_K^2,m_\pi^2,t,m_\pi^2) 
              ( 1/3 m_\pi^2 t - 1/8 t^2 )
   \nonumber\\&& + V_{24}(m_\eta^2,m_\eta^2,m_K^2,m_K^2,m_\pi^2,t,m_\pi^2) 
                (  - 1/3 m_\pi^2 t + 1/8 
         t^2 )
   \nonumber\\&& + V_{25}(m_\pi^2,m_\pi^2,m_\pi^2,m_\pi^2,m_\pi^2,t,m_\pi^2) 
            (  - 4/3 t )
   \nonumber\\&& + V_{25}(m_\pi^2,m_\pi^2,m_K^2,m_K^2,m_\pi^2,t,m_\pi^2)  
              ( 4 m_\pi^2 - 2 t )
   \nonumber\\&& + V_{25}(m_\pi^2,m_\eta^2,m_K^2,m_K^2,m_\pi^2,t,m_\pi^2)  
             (  - m_\pi^2 + 1/2 t )
   \nonumber\\&& + V_{25}(m_K^2,m_K^2,m_\pi^2,m_K^2,m_\pi^2,t,m_\pi^2)  
             ( 4 m_\pi^2 - 19/12 t )
   \nonumber\\&& + V_{25}(m_K^2,m_K^2,m_K^2,m_\eta^2,m_\pi^2,t,m_\pi^2) 
           ( 3 m_\pi^2 - 5/4 t )
   \nonumber\\&& + V_{25}(m_\eta^2,m_\pi^2,m_K^2,m_K^2,m_\pi^2,t,m_\pi^2) 
              (  - m_\pi^2 + 1/6 t )
   \nonumber\\&& + V_{26}(m_\pi^2,m_\pi^2,m_\pi^2,m_\pi^2,m_\pi^2,t,m_\pi^2)  
            ( 4/3 m_\pi^2 t - 2/3 t^2 )
   \nonumber\\&& + V_{26}(m_\pi^2,m_\pi^2,m_K^2,m_K^2,m_\pi^2,t,m_\pi^2) 
              (  - 2 m_\pi^2 t + 4 m_\pi^4 )
   \nonumber\\&& + V_{26}(m_\pi^2,m_\eta^2,m_K^2,m_K^2,m_\pi^2,t,m_\pi^2) 
             ( 1/2 m_\pi^2 t - m_\pi^4 )
   \nonumber\\&& + V_{26}(m_K^2,m_K^2,m_\pi^2,m_K^2,m_\pi^2,t,m_\pi^2)  
              (  - 29/12 m_\pi^2 t + 4 
         m_\pi^4 + 5/24 t^2 )
   \nonumber\\&& + V_{26}(m_K^2,m_K^2,m_K^2,m_\eta^2,m_\pi^2,t,m_\pi^2) 
             (  - 7/4 m_\pi^2 t + 3 m_\pi^4
          + 1/8 t^2 )
   \nonumber\\&& + V_{26}(m_\eta^2,m_\pi^2,m_K^2,m_K^2,m_\pi^2,t,m_\pi^2)  
             ( 5/6 m_\pi^2 t - m_\pi^4 - 1/6
          t^2 )
   \nonumber\\&& + V_{27}(m_\pi^2,m_\pi^2,m_\pi^2,m_\pi^2,m_\pi^2,t,m_\pi^2) 
              (  - 4/3 t^2 )
   \nonumber\\&& + V_{27}(m_\pi^2,m_\pi^2,m_K^2,m_K^2,m_\pi^2,t,m_\pi^2)  
             (  - t^2 )
   \nonumber\\&& + V_{27}(m_\pi^2,m_\eta^2,m_K^2,m_K^2,m_\pi^2,t,m_\pi^2) 
              ( 1/4 t^2 )
   \nonumber\\&& + V_{27}(m_K^2,m_K^2,m_\pi^2,m_K^2,m_\pi^2,t,m_\pi^2)  
             (  - 7/12 t^2 )
   \nonumber\\&& + V_{27}(m_K^2,m_K^2,m_K^2,m_\eta^2,m_\pi^2,t,m_\pi^2) 
           (  - 1/2 t^2 )
   \nonumber\\&& + V_{27}(m_\eta^2,m_\pi^2,m_K^2,m_K^2,m_\pi^2,t,m_\pi^2) 
              (  - 1/12 t^2 )
   \nonumber\\&& + V_{28}(m_\pi^2,m_\pi^2,m_\pi^2,m_\pi^2,m_\pi^2,t,m_\pi^2) 
             ( 8/3 m_\pi^2 t - 4/3 t^2 )
   \nonumber\\&& + V_{28}(m_\pi^2,m_\pi^2,m_K^2,m_K^2,m_\pi^2,t,m_\pi^2)  
              (  - 1/2 t^2 )
   \nonumber\\&& + V_{28}(m_\pi^2,m_\eta^2,m_K^2,m_K^2,m_\pi^2,t,m_\pi^2)  
               ( 1/2 m_\pi^2 t )
   \nonumber\\&& + V_{28}(m_K^2,m_K^2,m_\pi^2,m_K^2,m_\pi^2,t,m_\pi^2) 
              (  - 5/6 m_\pi^2 t - 1/12 
         t^2 )
   \nonumber\\&& + V_{28}(m_K^2,m_K^2,m_K^2,m_\eta^2,m_\pi^2,t,m_\pi^2) 
               (  - 1/2 m_\pi^2 t - 1/8 
         t^2 )
   \nonumber\\&& + V_{28}(m_\eta^2,m_\pi^2,m_K^2,m_K^2,m_\pi^2,t,m_\pi^2)  
            ( 1/6 m_\pi^2 t - 1/12 t^2
          )
   \nonumber\\&& + V_{29}(m_\pi^2,m_\pi^2,m_\pi^2,m_\pi^2,m_\pi^2,t,m_\pi^2)  
              (  - 2/3 t^2 )
   \nonumber\\&& + V_{29}(m_\pi^2,m_\pi^2,m_K^2,m_K^2,m_\pi^2,t,m_\pi^2)  
            (  - 1/2 t^2 )
   \nonumber\\&& + V_{29}(m_\pi^2,m_\eta^2,m_K^2,m_K^2,m_\pi^2,t,m_\pi^2) 
              (  - 1/2 m_\pi^2 t + 1/4 
         t^2 )
   \nonumber\\&& + V_{29}(m_K^2,m_K^2,m_\pi^2,m_K^2,m_\pi^2,t,m_\pi^2)  
              (  - 7/24 t^2 )
   \nonumber\\&& + V_{29}(m_K^2,m_K^2,m_K^2,m_\eta^2,m_\pi^2,t,m_\pi^2)  
              (  - 1/4 t^2 )
   \nonumber\\&& + V_{29}(m_\eta^2,m_\pi^2,m_K^2,m_K^2,m_\pi^2,t,m_\pi^2) 
            ( 1/2 m_\pi^2 t - 1/6 t^2 )
   \nonumber\\&& + V_{210}(m_\pi^2,m_\pi^2,m_\pi^2,m_\pi^2,m_\pi^2,t,m_\pi^2)
             ( 8 m_\pi^2 - 4 t )
   \nonumber\\&& + V_{210}(m_\pi^2,m_\pi^2,m_K^2,m_K^2,m_\pi^2,t,m_\pi^2) 
              ( 4 m_\pi^2 - 2 t )
   \nonumber\\&& + V_{210}(m_K^2,m_K^2,m_\pi^2,m_K^2,m_\pi^2,t,m_\pi^2) 
               ( 3/2 m_\pi^2 - 3/4 t )
   \nonumber\\&& + V_{210}(m_K^2,m_K^2,m_K^2,m_\eta^2,m_\pi^2,t,m_\pi^2)  
            ( 3/2 m_\pi^2 - 3/4 t )
   \nonumber\\&& + V_{211}(m_\pi^2,m_\pi^2,m_\pi^2,m_\pi^2,m_\pi^2,t,m_\pi^2)  
           (  - 4 m_\pi^2 t + 8 m_\pi^4 )
   \nonumber\\&& + V_{211}(m_\pi^2,m_\pi^2,m_K^2,m_K^2,m_\pi^2,t,m_\pi^2) 
           (  - 2 m_\pi^2 t + 4 m_\pi^4 )
   \nonumber\\&& + V_{211}(m_K^2,m_K^2,m_\pi^2,m_K^2,m_\pi^2,t,m_\pi^2) 
             (  - 3/4 m_\pi^2 t + 3/2 
        m_\pi^4 )
   \nonumber\\&& + V_{211}(m_K^2,m_K^2,m_K^2,m_\eta^2,m_\pi^2,t,m_\pi^2)  
              (  - 3/4 m_\pi^2 t + 3/2 
        m_\pi^4 )
   \nonumber\\&& + V_{212}(m_\pi^2,m_\pi^2,m_\pi^2,m_\pi^2,m_\pi^2,t,m_\pi^2) 
                   (  - 2 t^2 )
   \nonumber\\&& + V_{212}(m_\pi^2,m_\pi^2,m_K^2,m_K^2,m_\pi^2,t,m_\pi^2) 
              (  - t^2 )
   \nonumber\\&& + V_{212}(m_K^2,m_K^2,m_\pi^2,m_K^2,m_\pi^2,t,m_\pi^2)  
           (  - 3/8 t^2 )
   \nonumber\\&& + V_{212}(m_K^2,m_K^2,m_K^2,m_\eta^2,m_\pi^2,t,m_\pi^2)  
              (  - 3/8 t^2 )
   \nonumber\\&& + V_{213}(m_\pi^2,m_\pi^2,m_\pi^2,m_\pi^2,m_\pi^2,t,m_\pi^2) 
               (  - 2 t^2 )
   \nonumber\\&& + V_{213}(m_\pi^2,m_\pi^2,m_K^2,m_K^2,m_\pi^2,t,m_\pi^2)  
             (  - t^2 )
   \nonumber\\&& + V_{213}(m_K^2,m_K^2,m_\pi^2,m_K^2,m_\pi^2,t,m_\pi^2)  
               (  - 3/8 t^2 )
   \nonumber\\&& + V_{213}(m_K^2,m_K^2,m_K^2,m_\eta^2,m_\pi^2,t,m_\pi^2)  
                (  - 3/8 t^2 )\,.
\ea

\ba
  F_{S\mathbf{B}}^\pi(t) &=&
       + \frac{1}{16\pi^2}   ( 
  - 11/81 m_\pi^2 m_K^2 /m_\eta^2 \overline{A}(m_\eta^2) 
- 5/108 m_\pi^2 \overline{A}(m_\eta^2) + 5/6 m_\pi^2 
         \overline{A}(m_\pi^2) 
\nonumber\\&& ~~~~~~
- 1/4 m_\pi^2 \overline{A}(m_K^2)
 + 7/324 m_\pi^4 /m_\eta^2 \overline{A}(m_\eta^2)
 - 5/54 m_K^2 \overline{A}(m_\eta^2) + 
         m_K^2 \overline{A}(m_\pi^2) 
\nonumber\\&& ~~~~~~
+ m_K^2 \overline{A}(m_K^2)
 + 16/81 m_K^4 /m_\eta^2 \overline{A}(m_\eta^2) )
 \nonumber\\&&
      + \left(\frac{1}{16\pi^2}\right)^2 
          (  - 55/162 m_\pi^2 m_K^2\pi^2 - 1015/432 m_\pi^2 m_K^2 
            + 23/108 m_\pi^2 
        \pi^2 t
\nonumber\\&& ~~~~~~
 + 23/18 m_\pi^2 t - 1171/1296 m_\pi^4\pi^2 
              - 8783/1728 m_\pi^4 + 
         17/216 m_K^2\pi^2 t
\nonumber\\&& ~~~~~~
 + 17/36 m_K^2 t - 11/36 m_K^4\pi^2 
            - 15/16 m_K^4
          )
\nonumber\\&&
       + \overline{B}(m_\pi^2,m_\pi^2,t)^2
   (  - 1/2 m_\pi^2 t + 1/4 m_\pi^4 + 2/9 t^2
          )
\nonumber\\&&
       + \overline{B}(m_\pi^2,m_\pi^2,t) \overline{B}(m_K^2,m_K^2,t)
   (  - 1/24 m_\pi^2 t
          + 7/72 t^2 )
\nonumber\\&&
       + \overline{B}(m_\pi^2,m_\pi^2,t) \overline{B}(m_\eta^2,m_\eta^2,t)
   ( 1/27 m_\pi^2 t + 1/
         18 m_\pi^4 )
\nonumber\\&&
       + \overline{B}(m_\pi^2,m_\pi^2,t) \overline{B}_1(m_\pi^2,m_\pi^2,t)
   (  - m_\pi^2 t + 10/9
          t^2 )
\nonumber\\&&
       + \overline{B}(m_\pi^2,m_\pi^2,t) \overline{B}_1(m_K^2,m_K^2,t)
   (  - 1/12 m_\pi^2 t
          + 17/72 t^2 )
   \nonumber\\&& + \overline{B}(m_\pi^2,m_\pi^2,t)
 \overline{B}_{21}(m_\pi^2,m_\pi^2,t)   ( 4/9 t^2 )
  + \overline{B}(m_\pi^2,m_\pi^2,t) 
\overline{B}_{21}(m_K^2,m_K^2,t)   ( 1/12 t^2 )
   \nonumber\\&& + \overline{B}(m_\pi^2,m_\pi^2,t)
 \overline{B}_{22}(m_\pi^2,m_\pi^2,t)   ( 4/9 t )
 + \overline{B}(m_\pi^2,m_\pi^2,t)
 \overline{B}_{22}(m_K^2,m_K^2,t)   ( 1/12 t )
   \nonumber\\&& + \overline{B}(m_\pi^2,m_\pi^2,t) 
  (  - 1/12 m_\pi^2 \overline{A}(m_\eta^2)
    - 15/4 m_\pi^2 \overline{A}(m_\pi^2) - 2   m_\pi^2 \overline{A}(m_K^2)
\nonumber\\&& ~~~~~~
 + 1/9 t \overline{A}(m_\eta^2) + t \overline{A}(m_\pi^2)
 + 35/72 t \overline{A}(m_K^2) )
   \nonumber\\&& + \overline{B}(m_\pi^2,m_\pi^2,0)  
 (  - 1/9 m_\pi^2 \overline{A}(m_\eta^2) + 1/3 m_\pi^2 \overline{A}(m_\pi^2) )
   \nonumber\\&& + \overline{B}(m_K^2,m_K^2,t)^2   ( 1/24 t^2 )
   \nonumber\\&& + \overline{B}(m_K^2,m_K^2,t) 
\overline{B}(m_\eta^2,m_\eta^2,t)   (  - 1/9 m_\pi^2 m_K^2 + 
         1/24 m_\pi^2 t - 1/27 m_K^2 t + 1/72 t^2 )
   \nonumber\\&& + \overline{B}(m_K^2,m_K^2,t) 
\overline{B}_1(m_\pi^2,m_\pi^2,t)   (  - 1/12 m_\pi^2 t
          + 1/4 t^2 )
   \nonumber\\&& + \overline{B}(m_K^2,m_K^2,t)
 \overline{B}_1(m_K^2,m_K^2,t)   ( 5/24 t^2 )
   \nonumber\\&& + \overline{B}(m_K^2,m_K^2,t)
 \overline{B}_1(m_\eta^2,m_\eta^2,t)   ( 1/12 m_\pi^2 t + 1/
         36 t^2 )
   \nonumber\\&& + \overline{B}(m_K^2,m_K^2,t)
 \overline{B}_{21}(m_\pi^2,m_\pi^2,t)   ( 1/9 t^2 )
   \nonumber\\&& + \overline{B}(m_K^2,m_K^2,t)
 \overline{B}_{21}(m_K^2,m_K^2,t)   ( 1/12 t^2 )
   \nonumber\\&& + \overline{B}(m_K^2,m_K^2,t)
 \overline{B}_{22}(m_\pi^2,m_\pi^2,t)   ( 1/9 t )
   \nonumber\\&& + \overline{B}(m_K^2,m_K^2,t)
 \overline{B}_{22}(m_K^2,m_K^2,t)   ( 1/12 t )
   \nonumber\\&& + \overline{B}(m_K^2,m_K^2,t)
   (  - 1/4 m_\pi^2 \overline{A}(m_\eta^2)
 - 3/4 m_\pi^2 \overline{A}(m_\pi^2) - 1/2 
         m_\pi^2 \overline{A}(m_K^2)
\nonumber\\&& ~~~~~~
 - 53/360 t \overline{A}(m_\eta^2) + 35/72 t \overline{A}(m_\pi^2)
 + 17/120 t \overline{A}(m_K^2) )
   \nonumber\\&& + \overline{B}(m_K^2,m_K^2,0)  
 ( 1/18 m_K^2 \overline{A}(m_\eta^2) )
   \nonumber\\&& + \overline{B}(m_\eta^2,m_\eta^2,t)^2 
  ( 4/81 m_\pi^2 m_K^2 - 7/324 m_\pi^4 )
   \nonumber\\&& + \overline{B}(m_\eta^2,m_\eta^2,t)
 \overline{B}_1(m_\pi^2,m_\pi^2,t)   ( 1/27 m_\pi^2 t )
   \nonumber\\&& + \overline{B}(m_\eta^2,m_\eta^2,t)
 \overline{B}_1(m_K^2,m_K^2,t)   ( 1/12 m_\pi^2 t - 1/
         27 m_K^2 t + 1/24 t^2 )
   \nonumber\\&& + \overline{B}(m_\eta^2,m_\eta^2,t) 
\overline{B}_{21}(m_K^2,m_K^2,t)   ( 1/36 t^2 )
   \nonumber\\&& + \overline{B}(m_\eta^2,m_\eta^2,t) 
\overline{B}_{22}(m_K^2,m_K^2,t)   ( 1/36 t )
   \nonumber\\&& + \overline{B}(m_\eta^2,m_\eta^2,t) 
  ( 1/108 m_\pi^2 \overline{A}(m_\eta^2) 
+ 7/36 m_\pi^2 \overline{A}(m_\pi^2) - 2/9 
         m_\pi^2 \overline{A}(m_K^2)
   \nonumber\\&& ~~~~~~        + 13/360 t \overline{A}(m_K^2) )
   \nonumber\\&& + \overline{B}^\epsilon(m_\pi^2,m_\pi^2,t) 
\frac{1}{16\pi^2}   (  - 2 m_\pi^2 m_K^2 + 14/9 m_\pi^2 t - 91/18 
         m_\pi^4 + 19/36 m_K^2 t
\nonumber\\&& ~~~~~~
  + 97/72 t^2 )
   \nonumber\\&& + \overline{B}^\epsilon(m_K^2,m_K^2,t)
\frac{1}{16\pi^2}  
 (  - m_\pi^2 m_K^2 + 103/180 m_\pi^2 t - 2/3 
         m_\pi^4 + 14/45 m_K^2 t
\nonumber\\&& ~~~~~~
 + 35/72 t^2 )
   \nonumber\\&& + \overline{B}^\epsilon(m_\eta^2,m_\eta^2,t) 
\frac{1}{16\pi^2}   ( 1/6 m_\pi^2 t - 5/54 m_\pi^4 + 1/20 m_K^2 
         t + 1/24 t^2 )
   \nonumber\\&& + \overline{B}_1^\epsilon(m_\pi^2,m_\pi^2,t)
 \frac{1}{16\pi^2}   ( 25/18 m_\pi^2 t + 17/18 m_K^2 t - 253/
         72 t^2 )
   \nonumber\\&& + \overline{B}_1^\epsilon(m_K^2,m_K^2,t)
 \frac{1}{16\pi^2}   (  - 71/180 m_\pi^2 t + 17/45 m_K^2 t - 
         101/72 t^2 )
   \nonumber\\&& + \overline{B}_1^\epsilon(m_\eta^2,m_\eta^2,t)
 \frac{1}{16\pi^2}   (  - 1/6 m_\pi^2 t - 1/10 m_K^2 t - 5/24
          t^2 )
   \nonumber\\&& + \overline{B}_{21}^\epsilon(m_\pi^2,m_\pi^2,t)
 \frac{1}{16\pi^2}   ( 55/72 t^2 )
 + \overline{B}_{21}^\epsilon(m_K^2,m_K^2,t)
 \frac{1}{16\pi^2}   ( 13/36 t^2 )
   \nonumber\\&& + \overline{B}_{21}^\epsilon(m_\eta^2,m_\eta^2,t) 
\frac{1}{16\pi^2}   ( 1/8 t^2 )
 + \overline{B}_{22}^\epsilon(m_\pi^2,m_\pi^2,t) \frac{1}{16\pi^2} 
  (  - 15/2 m_\pi^2 + 95/36 t )
   \nonumber\\&& + \overline{B}_{22}^\epsilon(m_K^2,m_K^2,t) 
\frac{1}{16\pi^2}   (  - 3 m_\pi^2 + 10/9 t )
   \nonumber\\&& 
+ \overline{B}_{22}^\epsilon(m_\eta^2,m_\eta^2,t) \frac{1}{16\pi^2} 
  (  - 1/2 m_\pi^2 + 1/4 t )
+ \overline{B}_1(m_\pi^2,m_\pi^2,t)^2   ( 4/9 t^2 )
   \nonumber\\&& + \overline{B}_1(m_\pi^2,m_\pi^2,t) 
\overline{B}_1(m_K^2,m_K^2,t)   ( 7/36 t^2 )
   \nonumber\\&& + \overline{B}_1(m_\pi^2,m_\pi^2,t)  
 ( 1/9 t \overline{A}(m_\eta^2) + 5/9 t \overline{A}(m_\pi^2) + 17/18 t
         \overline{A}(m_K^2) )
   \nonumber\\&& + \overline{B}_1(m_K^2,m_K^2,t)^2   ( 1/12 t^2 )
+ \overline{B}_1(m_K^2,m_K^2,t) \overline{B}_1(m_\eta^2,m_\eta^2,t) 
  ( 1/36 t^2 )
   \nonumber\\&& + \overline{B}_1(m_K^2,m_K^2,t) 
  ( 23/180 t \overline{A}(m_\eta^2) - 1/12 t \overline{A}(m_\pi^2) + 2/15 
         t \overline{A}(m_K^2) )
   \nonumber\\&& + \overline{B}_1(m_\eta^2,m_\eta^2,t) 
  (  - 1/10 t \overline{A}(m_K^2) )
       - 1/2 m_\pi^{-2} m_K^2 \left(\overline{A}(m_\pi^2)\right)^2 
\nonumber\\&&
+ 23/18 m_\pi^{-2} t \left(\overline{A}(m_\pi^2)\right)^2 
- 4/3 m_\pi^2 m_K^{-2} 
         \left(\overline{A}(m_K^2)\right)^2
\nonumber\\&&
 + 11/162 m_\pi^2 m_K^2 /m_\eta^4 \left(\overline{A}(m_\eta^2)\right)^2 
- 35/144 m_\pi^2 /m_\eta^2 \left(\overline{A}(m_\eta^2)\right)^2
\nonumber\\&&
          - 7/648 m_\pi^4 /m_\eta^4 \left(\overline{A}(m_\eta^2)\right)^2 
+ 17/36 m_K^{-2} t \left(\overline{A}(m_K^2)\right)^2
\nonumber\\&&
  - 7/36 m_K^2 
         /m_\eta^2 \left(\overline{A}(m_\eta^2)\right)^2 
- 8/81 m_K^4 /m_\eta^4 \left(\overline{A}(m_\eta^2)\right)^2
\nonumber\\&&
 + 13/36 \overline{A}(m_\eta^2) \overline{A}(m_\pi^2) - 7/18
          \overline{A}(m_\eta^2) \overline{A}(m_K^2) 
+ 1/36 \left(\overline{A}(m_\eta^2)\right)^2
 - \overline{A}(m_\pi^2) \overline{A}(m_K^2)
\nonumber\\&&
 - 1037/144 \left(\overline{A}(m_\pi^2)\right)^2 - 7/4 
         \left(\overline{A}(m_K^2)\right)^2\,.
\ea

\ba
  F_{S\mathbf{L}}^\pi(t) &=&
       + \frac{1}{16\pi^2}   (  - 16/9 m_\pi^2 m_K^2 L_2^r 
    - 16/27 m_\pi^2 m_K^2 L_3^r + 12 m_\pi^2 m_K^2 L_4^r - 
         24 m_\pi^2 m_K^2 L_6^r
\nonumber\\&& ~~~
 + 4 m_\pi^4 L_1^r 
+ 74/9 m_\pi^4 L_2^r + 56/27 m_\pi^4 L_3^r + 16/
         3 m_\pi^4 L_4^r + 16/3 m_\pi^4 L_5^r
\nonumber\\&& ~~~
 - 32/3 m_\pi^4 L_6^r 
- 32/3 m_\pi^4 L_8^r 
+ 104/
         9 m_K^4 L_2^r + 86/27 m_K^4 L_3^r + 8/3 m_K^4 L_4^r 
\nonumber\\&& ~~~
+ 4/3 m_K^4 L_5^r
 - 16/3 
         m_K^4 L_6^r - 8/3 m_K^4 L_8^r )
   \nonumber\\&&
 + \overline{B}(m_\pi^2,m_\pi^2,t)   ( 8 m_\pi^2 m_K^2 L_4^r
 - 16 m_\pi^2 m_K^2 L_6^r - 24 m_\pi^2 
         L_1^r t - 8 m_\pi^2 L_2^r t - 12 m_\pi^2 L_3^r t
\nonumber\\&& ~~~
 - 12 m_\pi^2 L_4^r t - 26/3 m_\pi^2 
         L_5^r t + 32 m_\pi^2 L_6^r t + 64/3 m_\pi^2 L_8^r t 
+ 48 m_\pi^4 L_1^r
 + 16 m_\pi^4
          L_2^r
\nonumber\\&& ~~~
 + 24 m_\pi^4 L_3^r - 36 m_\pi^4 L_4^r - 16 m_\pi^4 L_5^r
 + 56 m_\pi^4 L_6^r + 24
          m_\pi^4 L_8^r - 32/3 m_K^2 L_4^r t
\nonumber\\ &&~~~ + 64/3 m_K^2 L_6^r t )
   \nonumber\\&& + \overline{B}(m_\pi^2,m_\pi^2,0) 
  ( 32/3 m_\pi^2 m_K^2 L_4^r - 64/3 m_\pi^2 m_K^2 L_6^r + 16/3 
         m_\pi^4 L_4^r + 16/3 m_\pi^4 L_5^r
\nonumber\\&& ~~~
 - 32/3 m_\pi^4 L_6^r - 32/3 m_\pi^4 L_8^r )
   \nonumber\\&& 
 + \overline{B}(m_K^2,m_K^2,t) 
  ( 32 m_\pi^2 m_K^2 L_1^r + 8 m_\pi^2 m_K^2 L_3^r - 32 m_\pi^2 
         m_K^2 L_4^r - 8 m_\pi^2 m_K^2 L_5^r
\nonumber\\&& ~~~
 + 32 m_\pi^2 m_K^2 L_6^r + 16 m_\pi^2 m_K^2 L_8^r - 4/3 m_\pi^2 
         L_4^r t + 8/3 m_\pi^2 L_6^r t - 16 m_K^2 L_1^r t
\nonumber\\&& ~~~
 - 4 m_K^2 L_3^r t
 - 8/3 m_K^2 
         L_5^r t + 16 m_K^2 L_6^r t + 16/3 m_K^2 L_8^r t )
   \nonumber\\&&
 + \overline{B}(m_K^2,m_K^2,0)   ( 4/3 m_\pi^2 m_K^2 L_4^r
 - 8/3 m_\pi^2 m_K^2 L_6^r + 8/3 
         m_K^4 L_4^r + 4/3 m_K^4 L_5^r 
\nonumber\\&&  ~~~
- 16/3 m_K^4 L_6^r - 8/3 m_K^4 L_8^r )
  \nonumber\\&&
 + \overline{B}(m_\eta^2,m_\eta^2,t)  
 ( 64/9 m_\pi^2 m_K^2 L_1^r + 32/27 m_\pi^2 m_K^2 L_3^r - 
         104/9 m_\pi^2 m_K^2 L_4^r
\nonumber\\&& ~~~
 - 64/27 m_\pi^2 m_K^2 L_5^r
 + 16 m_\pi^2 m_K^2 L_6^r - 128/9 m_\pi^2 
         m_K^2 L_7^r + 8/9 m_\pi^2 L_1^r t
\nonumber\\&& ~~~
 + 4/27 m_\pi^2 L_3^r t 
- 4/9 m_\pi^2 L_4^r t - 2/9 
         m_\pi^2 L_5^r t - 16/9 m_\pi^4 L_1^r
\nonumber\\&& ~~~
 - 8/27 m_\pi^4 L_3^r
 + 20/9 m_\pi^4 L_4^r 
+ 16/27 m_\pi^4 L_5^r - 8/3 m_\pi^4 L_6^r + 128/9 m_\pi^4 L_7^r
\nonumber\\&& ~~~
 + 40/9 m_\pi^4 L_8^r - 32/
         9 m_K^2 L_1^r t
 - 16/27 m_K^2 L_3^r t + 16/9 m_K^2 L_4^r t )
  \nonumber\\&&
 + \overline{B}_1(m_\pi^2,m_\pi^2,t)   (  - 48 m_\pi^2 L_1^r t
 - 16 m_\pi^2 L_2^r t - 24 
         m_\pi^2 L_3^r t - 8/3 m_\pi^2 L_5^r t + 32 m_\pi^2 L_6^r t
\nonumber\\&& ~~~
 + 64/3 m_\pi^2 L_8^r t - 
         32/3 m_K^2 L_4^r t + 64/3 m_K^2 L_6^r t + 32 L_1^r t^2 + 24 L_2^r t^2
\nonumber\\&& ~~~
 +       16 L_3^r t^2 + 32/3 L_4^r t^2 + 16/3 L_5^r t^2 )
  \nonumber\\&& 
+ \overline{B}_1(m_K^2,m_K^2,t)  
 (  - 32 m_\pi^2 L_1^r t - 8 m_\pi^2 L_3^r t + 44/3 
         m_\pi^2 L_4^r t + 4 m_\pi^2 L_5^r t 
\nonumber\\&& ~~~
+ 8/3 m_\pi^2 L_6^r t
 - 8 m_K^2 L_4^r t - 8/3 
         m_K^2 L_5^r t + 16 m_K^2 L_6^r t + 16/3 m_K^2 L_8^r t 
+ 16 L_1^r t^2
\nonumber\\&& ~~~
 + 8   L_2^r t^2
 + 6 L_3^r t^2 + 16/3 L_4^r t^2 + 4/3 L_5^r t^2 )
  \nonumber\\&&
 + \overline{B}_1(m_\eta^2,m_\eta^2,t) 
  (  - 16/3 m_\pi^2 L_1^r t - 8/9 m_\pi^2 L_3^r t + 
         16/3 m_\pi^2 L_4^r t + 8/9 m_\pi^2 L_5^r t
\nonumber\\&& ~~~
 + 8/3 L_1^r t^2 + 4/3 L_2^r t^2 + 
         8/9 L_3^r t^2 )
  \nonumber\\&&
 + \overline{B}_{21}(m_\pi^2,m_\pi^2,t) 
  (  - 8 L_1^r t^2 - 16 L_2^r t^2 - 4 L_3^r 
         t^2 + 32/3 L_4^r t^2 + 16/3 L_5^r t^2 )
  \nonumber\\&&
 + \overline{B}_{21}(m_K^2,m_K^2,t)  
 (  - 8 L_2^r t^2 - 2 L_3^r t^2 + 16/3 L_4^r 
         t^2 + 4/3 L_5^r t^2 )
  \nonumber\\&&
 + \overline{B}_{21}(m_\eta^2,m_\eta^2,t) 
  (  - 4/3 L_2^r t^2 - 4/9 L_3^r t^2 )
  \nonumber\\&&
 + \overline{B}_{22}(m_\pi^2,m_\pi^2,t)
   ( 32 m_\pi^2 L_1^r + 64 m_\pi^2 L_2^r + 16 m_\pi^2 L_3^r - 
         16 L_1^r t - 32 L_2^r t - 8 L_3^r t 
\nonumber\\&& ~~~
+ 32/3 L_4^r t + 16/3 L_5^r t )
   \nonumber\\&& + \overline{B}_{22}(m_K^2,m_K^2,t) 
  ( 32 m_\pi^2 L_2^r + 8 m_\pi^2 L_3^r - 16 L_2^r t - 4 
         L_3^r t + 16/3 L_4^r t + 4/3 L_5^r t )
   \nonumber\\&& + \overline{B}_{22}(m_\eta^2,m_\eta^2,t) 
  ( 16/3 m_\pi^2 L_2^r + 16/9 m_\pi^2 L_3^r - 8/3 L_2^r 
         t - 8/9 L_3^r t )
\nonumber\\&&
       - 128 m_\pi^2 m_K^2 L_4^r L_5^r + 1536 m_\pi^2 m_K^2 L_4^r L_6^r
 + 512 m_\pi^2 m_K^2 L_4^r L_8^r - 
         256 m_\pi^2 m_K^2 (L_4^r)^2 
\nonumber\\&&
+ 256 m_\pi^2 m_K^2 L_5^r L_6^r - 1024 m_\pi^2 m_K^2 L_6^r L_8^r - 2048
          m_\pi^2 m_K^2 (L_6^r)^2 + 104 m_\pi^2 L_1^r \overline{A}(m_\pi^2)
\nonumber\\&&
 + 32 m_\pi^2 L_1^r \overline{A}(m_K^2) - 4/3 m_\pi^2 L_2^r 
         \overline{A}(m_\eta^2) + 48 m_\pi^2 L_2^r \overline{A}(m_\pi^2)
 - 4/9 m_\pi^2 L_3^r \overline{A}(m_\eta^2)
\nonumber\\&&
 + 52 m_\pi^2 L_3^r \overline{A}(m_\pi^2) + 8 
         m_\pi^2 L_3^r \overline{A}(m_K^2) 
- 8/3 m_\pi^2 L_4^r \overline{A}(m_\eta^2) 
- 424/3 m_\pi^2 L_4^r \overline{A}(m_\pi^2)
\nonumber\\&&
 - 172/3 m_\pi^2 
         L_4^r \overline{A}(m_K^2) - 4/3 m_\pi^2 L_5^r \overline{A}(m_\eta^2)
 - 244/3 m_\pi^2 L_5^r \overline{A}(m_\pi^2)
 - 24 m_\pi^2 L_5^r 
         \overline{A}(m_K^2)
\nonumber\\&&
 + 800/3 m_\pi^2 L_6^r \overline{A}(m_\pi^2)
 + 248/3 m_\pi^2 L_6^r \overline{A}(m_K^2) 
+ 32 m_\pi^2 L_7^r \overline{A}(m_\eta^2)
          + 16 m_\pi^2 L_8^r \overline{A}(m_\eta^2)
\nonumber\\&&
 + 464/3 m_\pi^2 L_8^r \overline{A}(m_\pi^2) 
+ 48 m_\pi^2 L_8^r \overline{A}(m_K^2) 
- 128          m_\pi^4 L_4^r L_5^r 
+ 512 m_\pi^4 L_4^r L_6^r 
\nonumber\\&&
+ 384 m_\pi^4 L_4^r L_8^r 
- 64 m_\pi^4          (L_4^r)^2 
+ 512 m_\pi^4 L_5^r L_6^r
 + 384 m_\pi^4 L_5^r L_8^r 
- 64 m_\pi^4 (L_5^r)^2
\nonumber\\&&
 - 1280          m_\pi^4 L_6^r L_8^r
 - 768 m_\pi^4 (L_6^r)^2 
- 512 m_\pi^4 (L_8^r)^2
 + 64/3 m_K^2 L_1^r          \overline{A}(m_\eta^2)
\nonumber\\&&
 + 64 m_K^2 L_1^r \overline{A}(m_K^2)
 + 16/3 m_K^2 L_2^r \overline{A}(m_\eta^2)
 + 16 m_K^2 L_2^r \overline{A}(m_K^2) + 16/
         3 m_K^2 L_3^r \overline{A}(m_\eta^2) 
\nonumber\\&&
+ 20 m_K^2 L_3^r \overline{A}(m_K^2) 
- 88/3 m_K^2 L_4^r \overline{A}(m_\eta^2) 
- 104/3 m_K^2          L_4^r \overline{A}(m_\pi^2) 
\nonumber\\&&
- 248/3 m_K^2 L_4^r \overline{A}(m_K^2)
 - 32/9 m_K^2 L_5^r \overline{A}(m_\eta^2)
 - 52/3 m_K^2 L_5^r          \overline{A}(m_K^2) 
\nonumber\\&&
+ 112/3 m_K^2 L_6^r \overline{A}(m_\eta^2) 
+ 208/3 m_K^2 L_6^r \overline{A}(m_\pi^2)
 + 304/3 m_K^2 L_6^r     \overline{A}(m_K^2)
\nonumber\\&&
 - 64/3 m_K^2 L_7^r \overline{A}(m_\eta^2)
 + 104/3 m_K^2 L_8^r \overline{A}(m_K^2)
 - 128 m_K^4 L_4^r L_5^r
          + 1024 m_K^4 L_4^r L_6^r
\nonumber\\&&
+ 256 m_K^4 L_4^r L_8^r 
- 256 m_K^4 (L_4^r)^2
 + 256    m_K^4 L_5^r L_6^r
 - 512 m_K^4 L_6^r L_8^r 
- 1024 m_K^4 (L_6^r)^2
 \nonumber\\&&
 - 8/3 L_1^r t          \overline{A}(m_\eta^2)
- 24 L_1^r t \overline{A}(m_\pi^2)
 - 16 L_1^r t \overline{A}(m_K^2)
 - 8 L_2^r t \overline{A}(m_\pi^2)
 - 4/9          L_3^r t \overline{A}(m_\eta^2)
 \nonumber\\&&
 - 12 L_3^r t \overline{A}(m_\pi^2)
 - 4 L_3^r t \overline{A}(m_K^2)
 + 4/3 L_4^r t      \overline{A}(m_\eta^2)
 + 20/3 L_4^r t \overline{A}(m_\pi^2)
 \nonumber\\&&
 + 16/3 L_4^r t \overline{A}(m_K^2) 
+ 2/3 L_5^r t \overline{A}(m_\eta^2)
 + 10/3 L_5^r t \overline{A}(m_\pi^2) 
+ 10/3 L_5^r t \overline{A}(m_K^2)
\,.
\ea


\begin{thebibliography}{99}

\bibitem{Weinberg}
S.~Weinberg,
Physica A {\bf 96} (1979) 327.

\bibitem{GL1}
J.~Gasser and H.~Leutwyler,
Annals Phys.\  {\bf 158} (1984) 142.

\bibitem{GL2}
J.~Gasser and H.~Leutwyler,
Nucl.\ Phys.\ B {\bf 250} (1985) 465.

\bibitem{chptlectures}
A.~Pich, Lectures at Les Houches Summer School in
Theoretical Physics, Session 68: Probing the Standard Model of Particle
Interactions, Les Houches, France, 28 Jul - 5 Sep 1997,
[hep-ph/9806303];\\
G.~Ecker,
Lectures given at Advanced School on Quantum Chromodynamics (QCD 2000),
Benasque, Huesca, Spain, 3-6 Jul 2000,
[hep-ph/0011026];\\
S.~Scherer,
hep-ph/0210398.

\bibitem{GL3}
J.~Gasser and H.~Leutwyler,
Nucl.\ Phys.\ B {\bf 250} (1985) 517.

\bibitem{largeNc}
G.~'t Hooft,
Nucl.\ Phys.\ B {\bf 72} (1974) 461;\\
A.~V.~Manohar,
hep-ph/9802419,
Les Houches Summer School in Theoretical Physics,
Session 68: Probing the Standard Model of Particle Interactions, Les Houches,
France, 28 Jul - 5 Sep 1997
F. David and R. Gupta eds.

\bibitem{Moussallam1}
B.~Moussallam,
Eur.\ Phys.\ J.\ C {\bf 14} (2000) 111
[hep-ph/9909292].

\bibitem{Stern1}
M.~Knecht, B.~Moussallam, J.~Stern and N.~H.~Fuchs,
Nucl.\ Phys.\ B {\bf 457} (1995) 513
[hep-ph/9507319].

\bibitem{pipi1}
J.~Bijnens, G.~Colangelo, G.~Ecker, J.~Gasser and M.~E.~Sainio,
Phys.\ Lett.\ B {\bf 374} (1996) 210
[hep-ph/9511397].

\bibitem{pipi2}
J.~Bijnens, G.~Colangelo, G.~Ecker, J.~Gasser and M.~E.~Sainio,
Nucl.\ Phys.\ B {\bf 508} (1997) 263
[Erratum-ibid.\ B {\bf 517} (1998) 639]
[hep-ph/9707291].

\bibitem{ACGL}
B.~Ananthanarayan, G.~Colangelo, J.~Gasser and H.~Leutwyler,
Phys.\ Rept.\  {\bf 353} (2001) 207
[hep-ph/0005297].

\bibitem{CGL1}
G.~Colangelo, J.~Gasser and H.~Leutwyler,
Nucl.\ Phys.\ B {\bf 603} (2001) 125
[hep-ph/0103088].

\bibitem{E865}
S.~Pislak {\it et al.}  [BNL-E865 Collaboration],
Phys.\ Rev.\ Lett.\  {\bf 87} (2001) 221801
[hep-ex/0106071],
Phys.\ Rev.\ D {\bf 67} (2003) 072004
[hep-ex/0301040].

\bibitem{CGL2}
G.~Colangelo, J.~Gasser and H.~Leutwyler,
Phys.\ Rev.\ Lett.\  {\bf 86} (2001) 5008
[hep-ph/0103063].

\bibitem{Stern2}
L.~Girlanda, M.~Knecht, B.~Moussallam and J.~Stern,
Phys.\ Lett.\ B {\bf 409} (1997) 461
[hep-ph/9703448].

\bibitem{Stern3}
S.~Descotes-Genon, N.~H.~Fuchs, L.~Girlanda and J.~Stern,
Eur.\ Phys.\ J.\ C {\bf 24} (2002) 469
[hep-ph/0112088].


\bibitem{Moussallam2}
B.~Moussallam,
JHEP {\bf 0008} (2000) 005
[hep-ph/0005245].

\bibitem{paramagnetic}
S.~Descotes-Genon, L.~Girlanda and J.~Stern,
JHEP {\bf 0001} (2000) 041
[hep-ph/9910537];\\
S.~Descotes-Genon and J.~Stern,
Phys.\ Lett.\ B {\bf 488} (2000) 274
[hep-ph/0007082];\\
S.~Descotes-Genon,
JHEP {\bf 0103} (2001) 002
[hep-ph/0012221];\\
S.~Descotes-Genon, L.~Girlanda and J.~Stern,
Eur.\ Phys.\ J.\ C {\bf 27} (2003) 115
[hep-ph/0207337].

\bibitem{BCT}
J.~Bijnens, G.~Colangelo and P.~Talavera,
JHEP {\bf 9805} (1998) 014
[hep-ph/9805389].

\bibitem{GM}
J.~Gasser and U.~G.~Meissner,
Nucl.\ Phys.\ B {\bf 357} (1991) 90.

\bibitem{ABT1}
G.~Amor\'os, J.~Bijnens and P.~Talavera,
Nucl.\ Phys.\ B {\bf 568} (2000) 319
[hep-ph/9907264].

\bibitem{GK}
E.~Golowich and J.~Kambor,
Phys.\ Rev.\ D {\bf 58} (1998) 036004
[hep-ph/9710214].

\bibitem{ABT2}
G.~Amor\'os, J.~Bijnens and P.~Talavera,
Phys.\ Lett.\ B {\bf 480} (2000) 71
[hep-ph/9912398].

\bibitem{ABT3}
G.~Amor\'os, J.~Bijnens and P.~Talavera,
Nucl.\ Phys.\ B {\bf 585} (2000) 293
[Erratum-ibid.\ B {\bf 598} (2001) 665]
[hep-ph/0003258].

\bibitem{ABT4}
G.~Amor\'os, J.~Bijnens and P.~Talavera,
Nucl.\ Phys.\ B {\bf 602} (2001) 87
[hep-ph/0101127].

\bibitem{BT2}
J.~Bijnens and P.~Talavera,
hep-ph/0303103.

\bibitem{Sirlin}
A. Sirlin, Ann. of Phys. 61 (1970) 294,
Phys.\ Rev.\ Lett.\  {\bf 43} (1979) 904.

\bibitem{BCE}
J.~Bijnens, G.~Colangelo and G.~Ecker,
JHEP {\bf 9902} (1999) 020
[hep-ph/9902437].

\bibitem{FS}
H.~W.~Fearing and S.~Scherer,
Phys.\ Rev.\ D {\bf 53} (1996) 315
[hep-ph/9408346].

\bibitem{BCE2}
J.~Bijnens, G.~Colangelo and G.~Ecker,
Annals Phys.\  {\bf 280} (2000) 100
[hep-ph/9907333].

\bibitem{BT1}
J.~Bijnens and P.~Talavera,
JHEP {\bf 0203} (2002) 046
[hep-ph/0203049].

\bibitem{Meissner1}
M.~Frink, B.~Kubis and U.~G.~Meissner,
Eur.\ Phys.\ J.\ C {\bf 25} (2002) 259
[hep-ph/0203193].

\bibitem{DGL}
J.~F.~Donoghue, J.~Gasser and H.~Leutwyler,
Nucl.\ Phys.\ B {\bf 343} (1990) 341.

\bibitem{Meissner2}
U.~G.~Meissner and J.~A.~Oller,
Nucl.\ Phys.\ A {\bf 679} (2001) 671
[hep-ph/0005253].

\bibitem{ABM}
B.~Ananthanarayan, P.~Buttiker and B.~Moussallam,
Eur.\ Phys.\ J.\ C {\bf 22} (2001) 133
[hep-ph/0106230].

\bibitem{Au}
K.~L.~Au, D.~Morgan and M.~R.~Pennington,
Phys.\ Rev.\ D {\bf 35} (1987) 1633.

\bibitem{KKL}
R.~Kami\'nski, L.~Le\'sniak and B.~Loiseau,
Phys.\ Lett.\ B {\bf 413} (1997) 130
[hep-ph/9707377].

\bibitem{JOP}
M.~Jamin, J.~A.~Oller and A.~Pich,
Nucl.\ Phys.\ B {\bf 622} (2002) 279
[hep-ph/0110193].

\bibitem{EGPR}
G.~Ecker, J.~Gasser, A.~Pich and E.~de Rafael,
Nucl.\ Phys.\ B {\bf 321} (1989) 311;\\
V.~Cirigliano, G.~Ecker, H.~Neufeld and A.~Pich,
JHEP {\bf 0306} (2003) 012
[hep-ph/0305311];\\
J.~Bijnens, E.~Gamiz, E.~Lipartia and J.~Prades,
JHEP {\bf 0304} (2003) 055
[hep-ph/0304222];\\
S.~Peris, M.~Perrottet and E.~de Rafael,
JHEP {\bf 9805} (1998) 011
[hep-ph/9805442].

\bibitem{FORM3}
J.~A.~Vermaseren,
math-ph/0010025.

\bibitem{Ghinculov}
A.~Ghinculov and J.J.~van der Bij,
Nucl.\ Phys.\ {\bf B436} (1995) 30
[hep-ph/9405418];\\
A.~Ghinculov and Y.~Yao,
Nucl.\ Phys.\ {\bf B516} (1998) 385
[hep-ph/9702266].


\end{thebibliography}
\end{document}